\documentclass[10pt,nofootinbib,preprintnumbers,superscriptaddress]{revtex4-1}

\usepackage{amssymb, amsfonts, amsmath, bm}
\usepackage{color}

\usepackage{mathrsfs}
\usepackage{enumerate}
\usepackage{amsfonts}
\usepackage{graphicx}

\newcommand{\bra}[1]{\left( #1 \right)}
\newcommand{\brb}[1]{\left[ #1 \right]}
\newcommand{\brc}[1]{\left\{ #1 \right\}}
\newcommand{\norm}[1]{\left| #1 \right|}
\newcommand{\MPl}{M_{\rm Pl}}
\newcommand{\rc}{r_{*c}}

\usepackage[
]{hyperref}
\hypersetup{%
 setpagesize=false,
 bookmarksnumbered=true,%
 bookmarksopen=true,%
 colorlinks=true,%
 linkcolor=blue,%
 citecolor=red}

\begin{document}
\title{
{\Large
Black hole perturbations in higher-order scalar-tensor theories: \texorpdfstring{\\}{}
initial value problem and dynamical stability
}}
\author{Keisuke Nakashi}
\affiliation{Department of Social Design Engineering, National Institute of Technology (KOSEN), Kochi College, 200-1 Monobe Otsu, Nankoku, Kochi, 783-8508, JAPAN}
\affiliation{Department of Physics, Rikkyo University, Toshima, Tokyo 171-8501, Japan}
\author{Masashi Kimura}
\affiliation{Department of Informatics and Electronics, Daiichi Institute of Technology, Tokyo 110-0005, Japan}
\affiliation{Department of Physics, Rikkyo University, Toshima, Tokyo 171-8501, Japan}
\author{Hayato Motohashi}
\affiliation{Division of Liberal Arts, Kogakuin University, 2665-1 Nakano-machi, Hachioji, Tokyo 192-0015, Japan}
\author{Kazufumi Takahashi}
\affiliation{Center for Gravitational Physics and Quantum Information, Yukawa Institute for Theoretical Physics, Kyoto University, 606-8502, Kyoto, Japan}
\date{\today}
\preprint{RUP-22-4, YITP-22-13}

\begin{abstract}
We propose a physically sensible formulation of initial value problem for black hole perturbations in higher-order scalar-tensor theories. 
As a first application, we study monopole perturbations around stealth Schwarzschild solutions in a shift- and reflection-symmetric subclass of degenerate higher-order scalar-tensor (DHOST) theories. 
In particular, we investigate the time evolution of the monopole perturbations by solving 
a two-dimensional wave equation and analyze the Vishveshwara's classical scattering experiment, i.e., the time evolution of a Gaussian wave packet. 
As a result, we confirm that stealth Schwarzschild solutions in the DHOST theory are dynamically stable against the monopole perturbations with the wavelength comparable or shorter than the size of the black hole horizon. 
We also find that the damped oscillations at the late time do not show up unlike the ringdown phase in the standard case of general relativity.
Moreover, we investigate the characteristic curves of the monopole perturbations as well as a static spherically symmetric solution with monopole hair. 

\end{abstract}
\maketitle

\section{Introduction}

Since the first detection of gravitational waves from a binary black hole merger~\cite{LIGOScientific:2016aoc}, 90 events have been reported by LIGO-Virgo-KAGRA Collaboration so far~\cite{LIGOScientific:2018mvr,LIGOScientific:2020ibl,LIGOScientific:2021usb,LIGOScientific:2021djp}. 
Furthermore, the first image of a black hole shadow of the M87 galactic center was also obtained by Event Horizon Telescope Collaboration~\cite{EventHorizonTelescope:2019dse}. 
The observational progress allow us to test gravity at a stronger-field and more dynamical regime than ever before, which motivates us to study modified gravity theories. 
According to the Lovelock's theorem~\cite{Lovelock:1971yv,Lovelock:1972vz}, general relativity with a cosmological constant is the only possible covariant metric theory that yields second-order Euler-Lagrange equations in four dimensions. Therefore, in order to go beyond general relativity, one needs to relax at least one of the assumptions of the Lovelock's theorem. One way is to introduce additional dynamical field(s) other than the metric. 
In particular, theories with a single scalar field are called scalar-tensor theories. 
Of course, there are many other ways of modification, but most of them are at least effectively introducing additional degree(s) of freedom. 
In this sense, scalar-tensor theories offer a practical framework as a representative of modified gravity. 
When constructing the action of scalar-tensor theories, there is an indicator for the (un)healthiness, i.e., the Ostrogradsky theorem~\cite{Woodard:2015zca,Motohashi:2020psc,Aoki:2020gfv}.
The theorem states that a system with nondegenerate higher derivatives has an unbounded Hamiltonian, and the associated degree(s) of freedom
are called Ostrogradsky ghosts.
In order to avoid this problem, we need to require that the higher-derivative terms are degenerate~\cite{Motohashi:2014opa,Langlois:2015cwa,Motohashi:2016ftl,Klein:2016aiq,Motohashi:2017eya,Motohashi:2018pxg} so that an appropriate linear combination of the Euler-Lagrange equations yields second-order differential equations.
In general, scalar-tensor theories with degenerate higher-derivative terms are called degenerate higher-order scalar-tensor (DHOST) theories.
A trivial class is the Horndeski theory~\cite{Horndeski:1974wa,Deffayet:2011gz,Kobayashi:2011nu}, for which the Euler-Lagrange equations themselves are of second order. 
There are many other classes of DHOST theories~\cite{Langlois:2015cwa,Crisostomi:2016czh,BenAchour:2016fzp,Takahashi:2017pje,Langlois:2018jdg,Takahashi:2021ttd}, which provide a general class of healthy scalar-tensor theories with higher derivatives.

So far, the observations about black holes are consistent with the Kerr metric, which is a unique rotating black hole solution in general relativity. 
However, these results do not necessarily mean that the only possible theory of gravity is general relativity. 
The reason is that there are modified gravity theories that allow the spacetime solutions of the same form as those in general relativity. 
In general, solutions in modified gravity that have the metric of the same form as in general relativity accompanied by nontrivial profile of the additional field(s) are called ``stealth'' solutions because there is no effect of modified gravity when we focus on the behaviors of test fields in such background spacetime. 
Stealth solutions in scalar-tensor theories have been actively studied in many works. 
In exploring solutions in scalar-tensor theories, it is practically useful to make a simple ansatz on the scalar field profile.
For a constant scalar field profile, a set of existence conditions for stealth solutions was derived in~\cite{Motohashi:2018wdq}.\footnote{The authors of \cite{Motohashi:2018wdq} investigated stealth black hole solutions in a quite general class of single-/multi-field scalar-tensor theories without imposing the shift symmetry. 
In the absence of the shift symmetry, a constant scalar field profile can be said to be nontrivial, while it is trivial in the presence of the shift symmetry.}
For a scalar field with a constant kinetic term, stealth black hole solutions have been obtained for the Schwarzschild--(anti-)de Sitter metric~\cite{Babichev:2013cya,Kobayashi:2014eva,Babichev:2016kdt,Babichev:2017guv,Babichev:2017lmw,BenAchour:2018dap,Motohashi:2019sen,Minamitsuji:2019shy,Khoury:2020aya} and for the Kerr-de Sitter metric~\cite{Babichev:2017guv,Charmousis:2019vnf}.
Most of these solutions were found in the so-called shift-symmetric scalar-tensor theories, in which static black holes can support time-dependent scalar hair (see Sec.~\ref{sec:theory}).\footnote{For a time-independent scalar field, static and spherically symmetric black hole solutions are linearly unstable in general~\cite{Minamitsuji:2018vuw,Minamitsuji:2022mlv}.}
Furthermore, existence conditions for stealth solutions in the presence of a general matter component were obtained in \cite{Takahashi:2020hso}.

Although the behaviors of test fields on stealth background are the same as those in general relativity, the behaviors of gravitational perturbations can be different. Therefore, investigating the perturbations about stealth solutions in modified gravity theories is important to test gravity with observations such as gravitational wave observations. 
Perturbations around stealth black hole solutions have been studied in \cite{Ogawa:2015pea,Takahashi:2015pad,Takahashi:2016dnv,Tretyakova:2017lyg,Babichev:2017lmw,Babichev:2018uiw,Minamitsuji:2018vuw,Takahashi:2019oxz,deRham:2019gha,Charmousis:2019fre,Khoury:2020aya,Tomikawa:2021pca,Langlois:2021aji,Langlois:2021xzq,Takahashi:2021bml}.  One of the issues predicted from the black hole perturbation theory and closely related to actual observations is the quasinormal modes. 
In general relativity, it is known that the quasinormal modes describe the exponentially damped oscillations of the black hole perturbation at the late time~\cite{Leaver:1986gd,Nollert:1992ifk,Andersson:1995zk,Andersson:1996cm,Berti:2006wq}. 
Furthermore, the ringdown waveform can be well described by the superposition of the quasinormal modes~\cite{Buonanno:2006ui}. 
The quasinormal modes have complex frequencies, whose imaginary part is the inverse of the damping time, where time is usually identified by the Killing time coordinate in the background black hole spacetime. 
In DHOST theories, there are some earlier works 
in which the quasinormal frequencies were calculated based on the perturbation equations of motion recast in well-known forms, e.g., in the form of a Schr\"odinger-type equation or first-order coupled equations~\cite{Tomikawa:2021pca,Langlois:2021aji,Langlois:2021xzq}.

When we transform the equations of motion into such forms, we need to introduce a new time coordinate in general. 
For instance, for static and spherically symmetric black holes with time-dependent scalar hair, the new time coordinate is a function of both the time and radius in the static coordinate system~\cite{Takahashi:2019oxz}.
In \cite{Tomikawa:2021pca,Langlois:2021aji,Langlois:2021xzq}, the properties of the new time coordinate were not studied in detail, and hence it remains unclear whether or not the new time coordinate is appropriate to describe the observables such as the damping time. 
In addition, there is another reason to investigate the properties of the new time coordinate. 
The equation of motion for the black hole perturbations can be often written in the form of a two-dimensional wave equation. 
When we solve it, we need to specify the initial surface of the Cauchy problem. 
Usually, we choose a spacelike hypersurface (e.g., a surface of constant Killing time coordinate) as an initial surface because it is known that initial value problems are well-posed when one imposes initial conditions on a spacelike hypersurface\footnote{Also, given initial conditions on a null hypersurface, we can solve the two-dimensional wave equation and obtain the time evolution of perturbations~\cite{Gundlach:1993tp}.} (see, e.g., \cite{Wald:1984rg} and references therein). 
However, it is nontrivial whether or not a surface of constant new time coordinate is spacelike.
Indeed, we show that a surface of constant new time coordinate is not spacelike in the DHOST theory considered in the present paper. Of course, from a mathematical point of view, one can solve the two-dimensional wave equation even if the initial surface is not spacelike.
Nevertheless, from a physical point of view, if the ``initial'' surface is not spacelike, the situation corresponds to imposing a boundary condition from the past timelike infinity to the future timelike infinity. 
The evolution of the perturbation imposing such a boundary condition would not be obtained as a solution of a physically sensible initial value problem.

A caveat is that perturbations around stealth solutions exhibit strong coupling in DHOST theories in general.
It was found that the sound speed for scalar waves vanishes in the asymptotic region~\cite{Babichev:2018uiw,Minamitsuji:2018vuw,deRham:2019gha,Motohashi:2019ymr,Takahashi:2021bml}, which signals strong coupling.
Also, in \cite{Takahashi:2021bml}, both odd- and even-parity perturbations about stealth Schwarzschild--(anti-)de Sitter solutions in the class of shift- and reflection-symmetric DHOST theories were investigated and it was pointed out that perturbations are either unstable or presumably strongly coupled.
These facts imply that the framework of DHOST theories per se cannot describe the perturbations about stealth solutions in a consistent manner.
For a consistent description, we need to introduce a higher-derivative term dubbed the ``scordatura'' term that weakly violates the degeneracy, which can cure the strong coupling problem keeping the mass scale of the Ostrogradsky ghost above the cutoff~\cite{Motohashi:2019ymr} (see Appendix~\ref{app:scordatura} for a brief review).
However, taking into account the scordatura term is technically involved because of the existence of the (harmless) Ostrogradsky mode.
Moreover, as mentioned earlier, a physically sensible formulation of initial value problem for black hole perturbations has not been established even in DHOST theories where the Ostrogradsky mode is absent.
Therefore, as a first step, it would be important to be equipped with a physically sensible formulation for DHOST theories, temporarily ignoring the strong coupling problem.
After establishing the formulation for DHOST theories, one can proceed to incorporate the scordatura term.
This is the reason why we focus on DHOST theories without the scordatura term in the present paper.

Having stated the importance as well as the subtlety of the time evolution of black hole perturbations in modified gravity,
in the present paper, we propose a physically sensible formulation of initial value problem for black hole perturbations in DHOST theories. 
As a first application, we obtain the time evolution of the monopole perturbations about stealth Schwarzschild solutions. 
Specifically, we analyze the Vishveshwara's classical scattering experiment~\cite{Vishveshwara:1970zz}, i.e., the time evolution of a Gaussian wave packet. We show that the stealth Schwarzschild solutions are dynamically stable against the monopole perturbations with the wavelength comparable or shorter than the size of the black hole horizon. 
We also show that the damped oscillation phase at the late time does not appear unlike the black hole perturbations in general relativity. 
Furthermore, we analyze several related issues, e.g., the characteristic curves\footnote{The characteristic curves are curves whose tangent vector field corresponds to the propagation directions of the fields at each point~\cite{courant1989}. In Appendix~\ref{app:algorithmcurves}, we discuss the characteristic curves in more detail.} for the monopole perturbations and the static and spherically symmetric perturbation. 
Our formulation would be applicable to a broader class of solutions and/or theories, including stealth black holes in the presence of the scordatura term.

The rest of the present paper is organized as follows. 
In Sec.~\ref{sec:theory}, we explain the framework of DHOST theories we consider and review stealth Schwarzschild solutions in DHOST theories. 
In Sec.~\ref{sec:perturbation}, we consider monopole perturbations about the stealth Schwarzschild solutions following the discussion in~\cite{Takahashi:2021bml}. 
We derive the quadratic Lagrangian for the monopole perturbations in terms of a single master variable. 
In Sec.~\ref{sec:waveequationandcurves}, 
we transform the equation of motion for the monopole perturbations into the form of a usual two-dimensional wave equation by introducing a new time coordinate, which we call $\tilde{t}$. 
We investigate the properties of the new time coordinate~$\tilde{t}$ as well as the characteristic curves of the monopole perturbations. 
In Sec.~\ref{sec:timeevolution}, we propose a physically sensible formulation of initial value problem in higher-order scalar-tensor theories. 
As an application, we study the time evolution of a Gaussian wave packet for the monopole perturbations. 
Section~\ref{sec:summary} is devoted to summary and discussions. 
The contents of Appendices are as follows: 
\begin{itemize}
    \item Appendix~\ref{app:scordatura}: 
    Strong coupling problem in perturbations around stealth solutions and the scordatura mechanism.
    \item Appendix~\ref{app:algorithmcurves}:
    Characteristic analysis of the monopole perturbations via an 
    algorithm proposed in~\cite{Motloch:2016msa}.
    \item Appendix~\ref{app:choiceinitialsurface}: Physical way to choose the initial surface for a two-dimensional wave equation.
    \item Appendix~\ref{app:method}: 
    Details of the numerical method and its convergence.
    \item Appendix~\ref{app:resultsother}: 
    Numerical results for cases that are not covered in the main text.
    \item Appendix~\ref{app:growingphase}:
    Analysis of the growing phase of the monopole perturbations with a toy model.
    \item Appendix~\ref{app:staticmode}: 
    Static and spherically symmetric perturbations around stealth background in DHOST theories.
\end{itemize}
Throughout the paper, we use the natural units in which $c=\hbar=1$.

\section{stealth Schwarzschild solutions in quadratic DHOST theory}
\label{sec:theory}

\subsection{Model}

We consider a class of quadratic DHOST theories~\cite{Langlois:2015cwa}, which is described by the following action: 
\begin{align}
  S = \int \mathrm{d}^{4}x \sqrt{-g}
      \brb{
        F_{0}(X)
      + F_{2}(X) R
      + \sum_{I=1}^{5} A_{I}(X) L_{I}^{(2)}
      },
      \label{eq:actionDHOST}
\end{align}
where the coupling functions~$F_0$, $F_2$, and $A_{I}$ $(I=1,\cdots ,5)$ are functions of $X\equiv \phi^{\mu} \phi_{\mu}$ and
\begin{align}
  L_{1}^{(2)} \equiv \phi^{\mu \nu} \phi_{\mu \nu},
  \quad
  L_{2}^{(2)} \equiv (\Box \phi)^{2},
  \quad
  L_{3}^{(2)} \equiv \phi^{\mu} \phi_{\mu \nu} \phi^{\nu} \Box \phi,
  \quad
  L_{4}^{(2)} \equiv \phi^{\mu} \phi_{\mu \nu} \phi^{\nu \lambda} \phi_{\lambda},
  \quad
  L_{5}^{(2)} \equiv (\phi^{\mu} \phi_{\mu \nu} \phi^{\nu})^{2},
\end{align}
with $\phi_{\mu} \equiv \nabla_{\mu} \phi$ and $\phi_{\mu \nu} \equiv \nabla_{\mu} \nabla_{\nu} \phi$. 
This action is invariant under the shift ($\phi \to \phi + \mathrm{const.}$) and the reflection ($\phi \to - \phi$) of the scalar field. 
For a generic choice of the coupling functions, there is a problem of Ostrogradsky ghost associated with higher-order Euler-Lagrange equations. 
On the other hand, this problem is absent if we impose the following degeneracy conditions:
\begin{align}
\label{eq:degeneracycondition}
\begin{split}
  A_{2} &= - A_{1} \neq - \frac{F_{2}}{X},
  \\
  A_{4} &= \frac{1}{8(F_{2} - X A_{1})^{2}}
           \Bigl\{
             4F_{2} \Bigl[ 3(A_{1} - 2F_{2X})^{2} - 2A_{3}F_{2}\Bigr]
           - A_{3}X^{2}(16A_{1}F_{2X} + A_{3}F_{2})
       \\
      &\hspace{2.8cm}
           + 4X
             (
             3A_{1} A_{3} F_{2} + 16 A_{1}^{2} F_{2X} - 16A_{1}F_{2X}^{2} - 4A_{1}^{3} +
               2A_{3}F_{2}F_{2X}
             )
           \Bigr\},
  \\
  A_{5} &= \frac{1}{8(F_{2}-XA_{1})^{2}}
           (2A_{1} - XA_{3} - 4F_{2X})
           \Bigl[
            A_{1}(2A_{1} + 3XA_{3} - 4F_{2X})
            - 4A_{3}F_{2}
           \Bigr],
\end{split}
\end{align}
where a subscript~$X$ denotes the derivative with respect to $X$. 
The class of theories described the action~\eqref{eq:actionDHOST} with~\eqref{eq:degeneracycondition} is called class Ia of quadratic DHOST theories~\cite{Langlois:2015cwa}.
The reason why we focus on this class is that all the other classes of quadratic DHOST theories are known to be either plagued by instabilities on a cosmological background or the modes corresponding to gravitational waves are absent.
For demonstration purposes, in the present paper, we consider a subclass of class Ia DHOST theories described by the following action:
\begin{align}
  S = \int \mathrm{d}^{4} x \sqrt{-g}
          \brb{
            F_{0}(X)
          + \frac{\MPl^{2}}{2} R
          + A_{3}(X) L_{3}^{(2)}
          + A_{4}(X) L_{4}^{(2)}
          + A_{5}(X) L_{5}^{(2)}
          },
          \label{eq:actionscordatura}
\end{align}
where the functional form of $A_4$ and $A_5$ is fixed by the degeneracy conditions~\eqref{eq:degeneracycondition} as
\begin{align}
  A_{4}(X) = - \frac{A_{3}}{4}
             \bra{
             \frac{4\MPl^{2} + X^{2}A_{3}}{\MPl^{2}}
             },
  \qquad
  A_{5}(X) = \frac{X A_{3}^{2}}{\MPl^{2}}.
\end{align}
Here, $\MPl=(8\pi G)^{-1/2}$ is the reduced Planck mass.

\subsection{Stealth Schwarzschild solutions}
\label{subsec:stealthsolution}

As the background spacetime, we focus on a stealth black hole solution.
A stealth black hole solution is such that the metric has the same form as the black hole solution in general relativity and the scalar field has a nontrivial configuration.
Recently, it was shown that any solution in general relativity in the presence of general matter components can be accommodated in higher-order scalar-tensor theories~\cite{Motohashi:2018wdq,Takahashi:2020hso}. 
If we focus on the Schwarzschild spacetime with a linearly time-dependent scalar field~$\phi=qt + \psi(r)$ having $X=-q^{2}$, the conditions for 
the theory~\eqref{eq:actionDHOST} to have the stealth Schwarzschild solutions are given by~\cite{Takahashi:2020hso}
\begin{align}
  F_{0}=0,
  \qquad
  F_{0X}=0,
\end{align}
which should hold at $X=-q^{2}$. 
Moreover, in DHOST theories satisfying \eqref{eq:degeneracycondition}, one can also show that the static and spherically symmetric black hole solution with such a scalar field profile is uniquely given by the Schwarzschild(-de Sitter) metric, with the value of $q$ (and the effective cosmological constant) fixed by the coupling functions in the action~\cite{Takahashi:2019oxz}.

For later convenience, we express the background Schwarzschild metric in terms of the Lema\^itre coordinates, where the metric has a Gaussian normal form: 
\begin{align}
  \bar g_{\mu \nu} \mathrm{d}x^{\mu} \mathrm{d}x^{\nu}
  = - \mathrm{d}\tau^{2} + (1 - A(r)) \mathrm{d}\rho^{2} + r^{2}(\tau, \rho) \gamma_{ab}\mathrm{d}x^{a}\mathrm{d}x^{b},
\end{align}
where the overbar denotes background quantity, $A(r)=1-r_{\mathrm{s}}/r$, $r_{\mathrm{s}}$ is the Schwarzschild radius, and $\gamma_{ab}$ is the metric on a unit two-dimensional sphere 
$\gamma_{ab}\mathrm{d}x^{a}\mathrm{d}x^{b}= 
\mathrm{d}\theta^2 + {\sin}^2\theta \mathrm{d}\varphi^2$. 
We note that the Latin indices~$a,b$ run over $\{\theta, \varphi\}$. 
The transformation law between the Lema\^itre coordinates~$\{ \tau, \rho, \theta, \varphi \} $ and the usual Schwarzschild coordinates~$\{t,r,\theta,\varphi\}$ is given by
\begin{align} \label{eq:trans}
  \mathrm{d} \tau = \mathrm{d}t + \frac{\sqrt{1-A(r)}}{A(r)} \mathrm{d}r,\qquad
  \mathrm{d} \rho = \mathrm{d}t + \frac{\mathrm{d}r}{A(r)\sqrt{1-A(r)}}.
\end{align}
These relations imply
\begin{align}
  &\tau = t + 2\sqrt{r_{\mathrm{s}}r} + r_{\mathrm{s}} \ln \left| \frac{\sqrt{r}-\sqrt{r_{\mathrm{s}}}}{\sqrt{r} + \sqrt{r_{\mathrm{s}}}} \right|, \qquad
  \rho = \tau + \frac{2}{3}r_{\mathrm{s}} \bra{ \frac{r}{r_{\mathrm{s}}} }^{3/2},
  \\
  &r(\tau,\rho) = \bra{ \frac{3}{2} (\rho - \tau) }^{2/3} r_{\mathrm{s}}^{1/3}.
\end{align}
\begin{figure}[t]
\includegraphics[width=160mm]{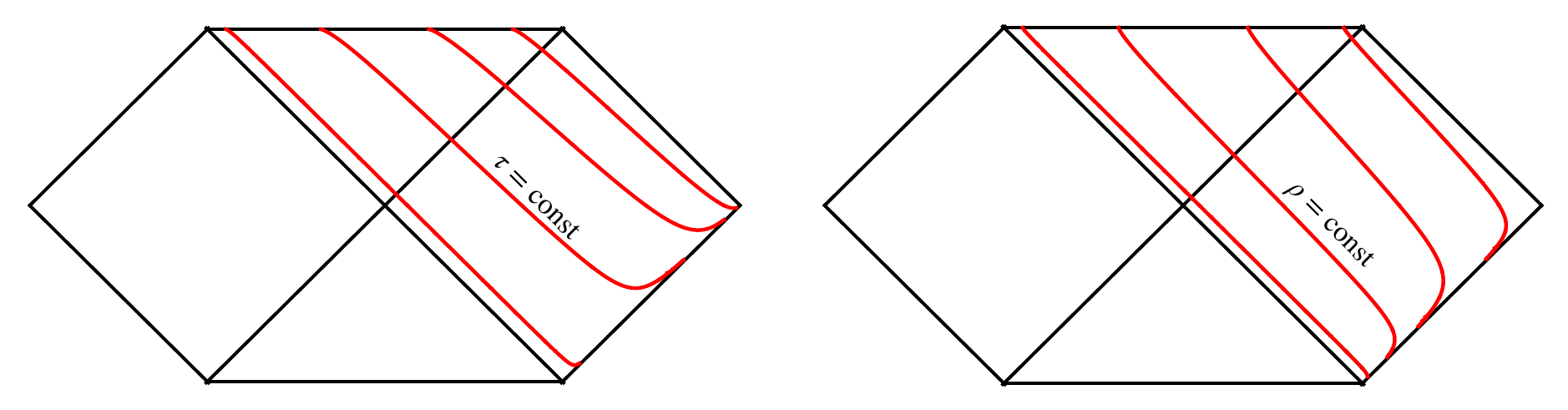}
\caption{The red curves depict $\tau=$ const.\ surfaces (left panel) and $\rho=$ const.\ surfaces (right panel). }
\label{fig:taurhoconst}
\end{figure}%
Figure~\ref{fig:taurhoconst} shows $\tau=$ const.~surfaces (left panel) and $\rho=$ const.~surfaces (right panel). 
Note that $\tau=$ const.~surfaces are spacelike everywhere, while $\rho=$ const.~surfaces are timelike everywhere.

In this coordinate system, the background scalar field profile satisfying $X=-q^2$ can be expressed as
\begin{align}
  \bar \phi = q \tau = q \bra{ t + 2\sqrt{r_{\mathrm{s}}r} + r_{\mathrm{s}} \ln \left| \frac{\sqrt{r}-\sqrt{r_{\mathrm{s}}}}{\sqrt{r} + \sqrt{r_{\mathrm{s}}}} \right| }.
  \label{eq:BGscalar}
\end{align}
Note that $\bar \phi$ is regular on the future event horizon. 
In fact, the scalar field behaves as $\bar \phi \sim q v + \mathrm{const}.$ in the vicinity of the future event horizon, where $v$ is the ingoing Eddington-Finkelstein coordinate defined by $v \equiv t + r + r_{\mathrm{s}} \ln |r/r_{\mathrm{s}}-1|$.\footnote{Precisely speaking, there are two branches of solution for $\bar{\phi}$.
The other branch is given by
    \begin{equation*}
    \bar{\phi}=q\bra{\tau-2\int \frac{\sqrt{1-A}}{A}{\rm d}r}
    = q \bra{ t - 2\sqrt{r_{\mathrm{s}}r} - r_{\mathrm{s}} \ln \left| \frac{\sqrt{r}-\sqrt{r_{\mathrm{s}}}}{\sqrt{r} + \sqrt{r_{\mathrm{s}}}} \right| },
    \end{equation*}
which is singular on the future event horizon.
In the present paper, we focus on the branch given by \eqref{eq:BGscalar}.}

\section{quadratic Lagrangian and master variable for monopole perturbations}
\label{sec:perturbation}

\subsection{Monopole perturbations}

We study linear perturbations about the stealth Schwarzschild solutions. 
It is known that the linear perturbations about a spherically symmetric spacetime can be decomposed into two modes: the odd- and even-parity modes. 
The odd- and even-parity perturbations for stealth Schwarzschild--(anti-)de Sitter solutions in DHOST theories were studied in~\cite{Takahashi:2019oxz,Khoury:2020aya, Tomikawa:2021pca,Takahashi:2021bml} and the conditions to avoid ghost and gradient instabilities were obtained. 
Since the background spacetime possesses the spherical symmetry, each component of perturbations can be expanded in terms of the spherical harmonics~$Y_{\ell m}(\theta, \varphi)$. 
In the present paper, we focus on the monopole perturbation ($\ell =m=0$) for simplicity, and hence the components of the metric perturbation~$\epsilon h_{\mu \nu} \equiv g_{\mu \nu} - \bar g_{\mu \nu}$, where $\epsilon$ is a small parameter that keeps track with perturbations, can be written as follows:
\begin{align}
  h_{\tau \tau} = H_{0}(\tau,\rho),\qquad
  h_{\tau \rho} =  H_{1}(\tau,\rho),\qquad
  h_{\rho \rho} = (1-A(r)) H_{2}(\tau,\rho),\qquad
  h_{ab}        = K(\tau,\rho) \gamma_{ab}.
\end{align}
For the scalar field, the perturbation~$\epsilon\delta\phi\equiv \phi-\bar{\phi}$ can be expressed as
\begin{align}
  \delta \phi = \delta \phi(\tau,\rho).
\end{align}

Considering an infinitesimal coordinate transformation~$x^{\mu} \to x^{\mu} + \epsilon \xi^{\mu}(\tau,\rho)$ ($\mu=\tau,\rho$), 
the gauge transformation law for the perturbation variables is given by
\begin{equation}
\begin{split}
  &H_{0} \to H_{0} + 2 \dot \xi^{\tau},\quad
  H_{1} \to H_{1} + \xi^{\tau \prime} - (1-A)\dot \xi^{\rho},\quad
  H_{2} \to H_{2} + \frac{A^{\prime}}{1-A}(\xi^{\rho}-\xi^{\tau}) - 2\xi^{\rho \prime},
  \\
  &K \to K - 2rr^{\prime}(\xi^{\rho}-\xi^{\tau}),
  \quad
  \delta \phi \to \delta \phi - q \xi^{\tau},
\end{split}
\end{equation}
where dots and primes denote the derivatives with respect to $\tau$ and $\rho$, respectively.
We fix the gauge degrees of freedom by choosing $K=\delta \phi = 0$,\footnote{Note that one cannot fix the gauge degrees of freedom so that $\delta \phi=0$ in the case where the constant $q$ vanishes. 
In the present paper, we assume that $q$ is a nonzero constant. } which is a complete gauge fixing in the sense that we can impose it at the action level without a loss of independent equations~\cite{Motohashi:2016prk}.
In this gauge, the metric perturbation~$h_{\mu \nu}$ and the scalar field are respectively given by 
\begin{align}
  h_{\mu \nu} \mathrm{d}x^{\mu} \mathrm{d}x^{\nu}
  &=
    H_{0}(\tau,\rho) \mathrm{d}\tau^{2}
  + 2 H_{1}(\tau,\rho) \mathrm{d}\tau \mathrm{d} \rho
  + (1-A(r))H_{2} (\tau,\rho) \mathrm{d}\rho^{2},
  \label{eq:perturbationmetric}
  \\
  \phi(\tau,\rho)
  &= \bar \phi.
  \label{eq:perscalar}
\end{align}

\subsection{Quadratic Lagrangian and master variable}
\label{ssec:monopole_master}

In this subsection, following \cite{Takahashi:2021bml}, 
we obtain the quadratic Lagrangian and the master variable for the monopole perturbations. 
Substituting the metric perturbation~\eqref{eq:perturbationmetric} into the action
of the DHOST theory~\eqref{eq:actionscordatura} and expanding it up to the second order in the small parameter $\epsilon$, we find the following quadratic Lagrangian:
\begin{align}
  {\cal L}^{(2)}
  =
   \;&a_{0} \dot H_{0}^{\prime} H_{1}
  + a_{1} \dot H_{0}^{2}
  + a_{2} \dot H_{0} \dot H_{2}
  + a_{3} H_{0}^{\prime 2}
  + b_{2} \dot H_{2} H_{0} \nonumber \\
  &+ b_{3} \dot H_{2} H_{1}
  + b_{6} H_{0}^{\prime} H_{1}
  + b_{7} H_{0}^{\prime} H_{2}
  + c_{1} H_{0}^{2}
  + c_{9} H_{2}^{2}.
\end{align}
Here, each coefficient is a function of $r$, whose expression can be found in \cite{Takahashi:2021bml}.
We introduce the following new variables: 
\begin{align}
  \tilde H_{1} \equiv H_{1} + (1-A)\bra{ \frac{a_{2}}{(1-A)b_{3}} H_{0}}^{\boldsymbol{\cdot}}, 
  \qquad
  \tilde H_{2} \equiv H_{2} + \frac{a_{0}}{b_{3}} H_{0}^{\prime},
\end{align}
to remove the terms proportional to $a_{0}$, $a_{1}$, $a_{2}$, and $a_{3}$.
In terms of these new variables, the quadratic Lagrangian is written as
\begin{align}
  {\cal L}^{(2)}
  =
    \tilde b_{2} \dot {\tilde{H}}_{2} H_{0}
  + \tilde b_{3} \dot {\tilde{H}}_{2} \tilde H_{1}
  + \tilde b_{6} H_{0}^{\prime} \tilde H_{1}
  + \tilde b_{7} H_{0}^{\prime} \tilde H_{2}
  + \tilde c_{1} H_{0}^{2}
  + \tilde c_{9} \tilde H_{2}^{2}.
\end{align}
The terms~$\dot{\tilde{H}}_{2} H_{0}$ and $\tilde H_{2}^{\prime} H_{0}$ can be removed by the following redefinition of $\tilde H_{1}$:
\begin{align}
  \zeta \equiv \tilde H_{1} + \frac{\tilde b_{2}}{\tilde b_{3}} H_{0} + \frac{\tilde b_{7}}{\tilde b_{6}} \tilde H_{2}.
\end{align}
By using this variable, the quadratic Lagrangian is written as
\begin{align}
  {\cal L}^{(2)}
  &=
    \zeta \bra{ \tilde b_{3} \dot{\tilde{H}}_{2} + \tilde b_{6} H_{0}^{\prime} }
  + \brb{
      \tilde c_{1} + \bra{ \frac{\tilde b_{2} \tilde b_{6}}{2 \tilde b_{3}} }^{\prime}\,
    } H_{0}^{2}
  + \brb{
      \tilde c_{9} + \bra{ \frac{\tilde b_{3} \tilde b_{7}}{2 \tilde b_{6}} }^{\boldsymbol{\cdot}}\, 
    } \tilde H_{2}^{2}, \notag
    \\
    &=
    \MPl^{2} r \zeta \brb{ \dot{\tilde{H}}_{2} + \bra{1 - \frac{3 q^{4} A_{3}}{2 \MPl^{2}}  } H_{0}^{\prime} }
    + 
    \frac{q^{4} F_{0XX} r^{2}\sqrt{1-A} }{2} H_{0}^{2}
    - 
    \frac{\MPl^{2} q^{4} A_{3} \sqrt{1-A} }{2\MPl^{2} -3q^{4}A_{3} }\tilde H_{2}^{2},
    \label{eq:2ndlagzetaH0tH2}
\end{align}
up to an overall numerical factor.

Let us make a comment on the general relativity limit, where $F_0(X)\to 0$ and $A_3(X)\to 0$ and hence the scalar field is absent. 
In this limit, the quadratic Lagrangian~\eqref{eq:2ndlagzetaH0tH2} takes the form
\begin{align}
    {\cal L}^{(2)} = 
    \MPl^2 r\zeta \bra{\dot{\tilde{H}}_{2} + H_{0}^{\prime} },
\end{align}
with
    \begin{equation}
    \tilde{H}_2=H_2, \qquad
    \zeta=H_1+\frac{1-A}{2}H_0+\frac{1}{2}H_2.
    \end{equation}
As it should be, this system does not have dynamical degrees of freedom, which corresponds to the well-known fact that the monopole perturbations are nondynamical in general relativity.
Note that the quadratic Lagrangian is now independent of $q$, reflecting the fact that the theory is independent of the scalar field. 
It should also be noted that this quadratic Lagrangian contains one gauge degree of freedom (see Appendix~C of \cite{Takahashi:2021bml}), which is nothing but the one we used to set $\delta\phi=0$ in the presence of the scalar field.

From the equations of motion for $H_{0}$ and $\tilde H_{2}$, 
we can express these variables in terms of $\zeta$ and its derivatives as
\begin{align}
  H_{0}
  =
  \frac{(2\MPl^{2} - 3q^{4}A_{3})(\sqrt{1-A}\zeta + r \zeta^{\prime})}{2q^{4}F_{0XX}r^{2} \sqrt{1-A}},
  \qquad
  \tilde H_{2}
  =
  \frac{(2\MPl^{2} - 3q^{4}A_{3})(\sqrt{1-A}\zeta - r \dot \zeta)}{2q^{4}A_{3} \sqrt{1-A}}.
  \label{eq:H0tH2DHOST}
\end{align}
Substituting these expressions back into the quadratic Lagrangian~\eqref{eq:2ndlagzetaH0tH2}, we obtain
\begin{align}
  {\cal L}^{(2)}
  =
    \frac{d_{1}}{2} \dot \zeta^{2}
  - \frac{d_{2}}{2} \zeta ^{\prime 2}
  - \frac{d_{3}}{2} \zeta ^{2},
  \label{eq:actionzeta}
\end{align}
where the coefficients~$d_{1}$, $d_{2}$, and $d_{3}$ are given by 
\begin{equation}
\begin{split}
  d_{1}
  & \equiv
  \frac{\MPl^{2} (2 \MPl^{2} - 3 q^{4} A_{3})}{2 q^{4} A_{3}} \frac{r^{2}}{\sqrt{1-A}}
  \equiv \frac{D_{1} r^{2}}{\sqrt{1-A}},\qquad
  d_{2}
  \equiv
  \frac{(2 \MPl^{2} - 3 q^{4} A_{3})^{2}}{4 q^{4} F_{0XX}\sqrt{1-A}}
  \equiv \frac{D_{2}}{\sqrt{1-A}}, \label{d1d2}
  \\
  d_{3}
  & \equiv
  \frac{2(1-A)}{r^{2}} d_{2},
\end{split}
\end{equation}
respectively. 
Here, the functions $A_{3}$ and $F_{0XX}$ are evaluated at $X=-q^{2}$. 
The necessary conditions for the absence of ghost and gradient instabilities are given by $d_{1}>0$ and $d_{2}>0$, which reduce to
\begin{align}
    D_{1}>0, \qquad
    D_{2}>0. 
\end{align}
From the quadratic Lagrangian~\eqref{eq:actionzeta}, we can read off the squared sound speed~$c_{s}^{2}$ of the monopole perturbations as
\begin{align}
  c_{s}^{2}
  =
  (1-A) \frac{d_{2}}{d_{1}}
  =
  \frac{D_{2}}{D_{1}} \frac{r_{\mathrm{s}}}{r^{3}}.
  \label{eq:soundspeedDHOST}
\end{align}
In the asymptotic flat region, the squared sound speed vanishes, which means that the scalar perturbation suffers from the strong coupling issue~\cite{Motohashi:2019ymr,Takahashi:2021bml}. 
This problem can be avoided by introducing a detuning term dubbed the scordatura term~\cite{Motohashi:2019ymr}, which weakly violates the degeneracy condition in a controlled manner. 
By introducing the scordatura term the Ostrogradsky ghost appears, but one can make the strong coupling scale sufficiently high so that the Ostrogradsky ghost shows up only above the cutoff scale of the effective field theory (EFT). 
We briefly review the scordatura mechanism in Appendix~\ref{app:scordatura}.

\section{Transformation into Wave equation and Characteristic curves}
\label{sec:waveequationandcurves}

In this section, we derive the equation of motion for the monopole perturbations in the form of a usual two-dimensional wave equation. 
In particular, we discuss a new time coordinate which is introduced to recast the equation of motion into the wave equation, and the characteristic curves of the monopole perturbations.

\subsection{Two-dimensional wave equation}

In order to obtain the equation of motion in the form of a two-dimensional wave equation, we go back from the Lema\^itre coordinates~$\{ \tau, \rho, \theta, \varphi \} $ to the Schwarzschild coordinates~$\{t,r,\theta,\varphi\}$ where all the metric components are static.
From Eq.~\eqref{eq:trans}, the inverse transformation law is given by
\begin{align}
  \mathrm{d} t = \frac{1}{A} \mathrm{d} \tau - \frac{1-A}{A} \mathrm{d} \rho, \qquad
  \mathrm{d} r = -\sqrt{1-A} \mathrm{d} \tau + \sqrt{1-A} \mathrm{d} \rho.
\end{align}
After performing the coordinate transformation, the quadratic Lagrangian is written as 
\begin{align}
  {\cal L}^{(2)} =   \frac{\hat d_{1}}{2} (\partial_{t} \zeta)^{2}
                       - \frac{\hat d_{2}}{2} (\partial_{r} \zeta)^{2}
                       - \frac{\hat d_{3}}{2} \zeta^{2}
                       +  \hat{d}_4(\partial_{t} \zeta)(\partial_{r} \zeta),
\label{eq:lagSchwarzschild}
\end{align}
where
\begin{align}
  \hat d_{1} \equiv \frac{d_{1} - (1-A)^{2} d_{2}}{A^{2} \sqrt{1-A}}, \qquad
  \hat d_{2} \equiv \sqrt{1-A} (d_{2} - d_{1}), \qquad
  \hat d_{3} \equiv \frac{d_{3}}{\sqrt{1-A}}, \qquad
  \hat d_{4} \equiv 
  \frac{(1-A) d_{2} - d_{1}}{A}.
\end{align}
Then, we remove the cross term~$\hat d_{4} (\partial_{t} \zeta) (\partial_{r} \zeta)$ by introducing a new time coordinate~$\tilde{t}$ such that
\begin{equation}
    {\rm d}\tilde{t}={\rm d}t+\frac{\hat{d}_4}{\hat{d}_2}{\rm d}r.
\end{equation}
We can write down $\tilde{t}$ explicitly as
\begin{align}
  \tilde t = t +
              \frac{2r^{3/2}}{3\sqrt{r_{\mathrm{s}}}}
              + 2 \sqrt{r_{\mathrm{s}}r}
              + \frac{r_{\mathrm{g}}^{3/2}}{\sqrt{r_{\mathrm{s}}}}\tan^{-1} \bra{ \sqrt{\frac{r}{r_{\mathrm{g}}}} }
              + \frac{r_{\mathrm{g}}^{3/2}}{2\sqrt{r_{\mathrm{s}}}} \ln \norm{ \frac{\sqrt{r}-\sqrt{r_{\mathrm{g}}}}{\sqrt{r}+\sqrt{r_{\mathrm{g}}}} }
              + r_{\mathrm{s}} \ln \norm{ \frac{\sqrt{r}-\sqrt{r_{\mathrm{s}}}}{\sqrt{r}+\sqrt{r_{\mathrm{s}}}} },
              \label{eq:newtimecoordinate}
\end{align}
where $r_{\mathrm{g}} \equiv \sqrt{D_{2}/D_{1}}$. 
Note that the use of $\tilde{t}$ does not spoil the static nature of the metric.
With this new time coordinate, the quadratic Lagrangian~\eqref{eq:lagSchwarzschild} takes the following diagonalized form: 
\begin{align}
  {\cal L}^{(2)} &=  \frac{\tilde d_{1}}{2} (\partial_{\tilde t} \zeta)^{2}
                    - \frac{\hat d_{2}}{2} (\partial_{r} \zeta)^{2}
                    - \frac{\hat d_{3}}{2} \zeta^{2},
                    \qquad
                    \tilde d_{1} \equiv \hat d_{1} + 
                    \frac{\hat d_{4}^{2}}{\hat d_{2}},
                    \label{eq:lagtildetr}
\end{align}
where the coefficients are given by
\begin{align}
  \tilde d_{1} = \frac{D_{1} r_{\mathrm{g}}^{2} r^{3}}{(r-r_{\mathrm{g}})(r+r_{\mathrm{g}})r_{\mathrm{s}}},
  \qquad
  \hat d_{2} = D_{1}(r-r_{\mathrm{g}})(r+r_{\mathrm{g}}),
  \qquad
  \hat d_{3} = -\frac{2 D_{1}r_{\mathrm{g}}^{2}}{r^{2}}.
\end{align}

Here, we discuss the characteristic horizon for the monopole perturbations. 
To do this, we focus on the kinetic part of the quadratic Lagrangian~\eqref{eq:lagtildetr} and define the effective metric~$Z_{IJ}$ with $I,J=(\tilde{t},r)$ so that 
\begin{align}
    {\cal L}_{\mathrm{kin}}^{(2)} = 
      \frac{\tilde{d}_{1}}{2} (\partial _{\tilde{t}}\zeta)^{2} 
    - \frac{\hat{d}_{2}}{2} (\partial _{r}\zeta)^{2}
    \equiv - \frac{1}{2} Z^{IJ} \partial_{I} \zeta \partial _{J} \zeta,
\end{align}
with $Z^{IJ}$ being the inverse of $Z_{IJ}$.
The components of the effective metric can be explicitly written as
\begin{align}
    Z_{IJ}\mathrm{d}x^{I} \mathrm{d}x^{J} = 
    - \frac{1}{\tilde{d}_{1}} \mathrm{d}\tilde{t}^{2}
    + \frac{1}{\hat{d}_{2}} \mathrm{d}r^{2}. 
    \label{eq:effectivemetrictr}
\end{align}
Since the spacetime described by $Z_{IJ}$ is static, 
the vector field~$(\partial_{\tilde t})^{\mu}$ is a timelike Killing vector field.
The Killing horizon for $(\partial_{\tilde t})^{\mu}$ is located at the radius where $Z_{\tilde{t}\tilde{t}}$ changes its sign.
Since $Z_{\tilde{t}\tilde{t}}=-1/\tilde{d}_1\propto r-r_{\rm g}$,
the Killing horizon is located at the radius~$r_{\mathrm{g}}=\sqrt{D_{2}/D_{1}}$.
In the rest of the present paper, we call $r=r_{\mathrm{g}}$ the monopole horizon.
In general, the two radii~$r_{\mathrm{s}}$ and $r_{\mathrm{g}}$ do not coincide with each other.\footnote{With the assumption that both the metric and the scalar field are stationary, the authors of \cite{Tanahashi:2017kgn,Benkel:2018qmh} studied conditions under which a Killing horizon of the metric also serves as a characteristic horizon for the other degree(s) of freedom.} 
The monopole horizon exists outside the Schwarzschild radius ($r_{\mathrm{g}} > r_{\mathrm{s}}$) if $D_{2}-D_{1}r_{\mathrm{s}}^{2}>0$, while the opposite ($r_{\mathrm{g}} < r_{\mathrm{s}}$) holds if $D_{2}-D_{1}r_{\mathrm{s}}^{2}<0$.
The radii of $r_{\mathrm{g}}$ and $r_{\mathrm{s}}$ coincide if $D_{2}-D_{1}r_{\mathrm{s}}^2=0$.

\begin{figure}[t]
\includegraphics[width=90mm]{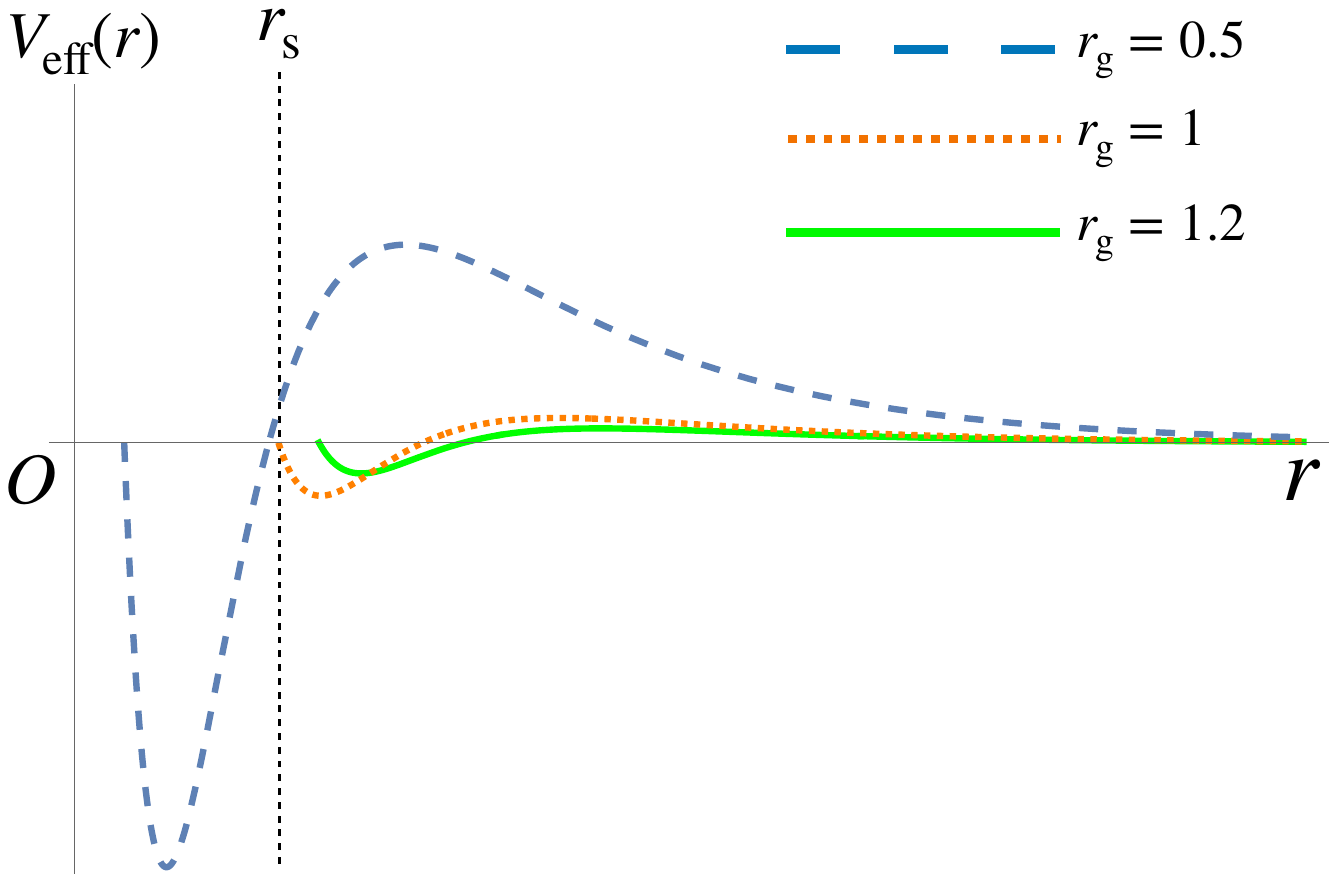}
\caption{Typical plots of the effective potential~$V_{\mathrm{eff}}$ for $r_{\mathrm{g}}=0.5$ (blue dashed curve), $r_{\mathrm{g}}=1$ (orange dotted curve), and $r_{\mathrm{g}}=1.2$ (green solid curve) in units of $r_{\rm s}$ (the position of $r=r_{\mathrm{s}}$ is depicted in a black dashed line).  
The left end point for each plot corresponds to $r=r_{\mathrm{g}}$. 
}
\label{fig:Veffplot}
\end{figure}

Finally, let us recast the equation of motion for the monopole perturbations into the form of a two-dimensional wave equation.
For this purpose, we focus on the region~$r>r_{\rm g}$ and introduce a generalized tortoise coordinate~$r_{*}$ and a new variable~$\Psi$ as follows:
\begin{align}
  r_{*} = \int \sqrt{\frac{\tilde d_{1}}{\hat d_{2}}} \mathrm{d} r, \label{eq:genetortoisecoordinate_int}
  \\
  \Psi = (\tilde d_{1} \hat d_{2})^{1/4} \zeta.
  \label{eq:Psizetarelation}
\end{align}
Since the integrand in the definition of $r_{*}$ is given by
 \begin{align}
   \sqrt{\frac{\tilde d_{1}}{\hat d_{2}}} = 
   \frac{r_{\mathrm{g}} r^{3/2}}{\sqrt{r_{\mathrm{s}}}(r- r_{\mathrm{g}})(r+ r_{\mathrm{g}})},
 \end{align}
the generalized tortoise coordinate $r_{*}$ can be expressed as
 \begin{align}
   r_{*}
   =
     2r_{\mathrm{g}}\sqrt{\frac{r}{r_{\mathrm{s}}}}
  - \frac{r_{\mathrm{g}}^{3/2}}{\sqrt{r_\mathrm{s}}}  \tan^{-1} \bra{ \sqrt{\frac{r}{r_{\mathrm{g}}}} }
   + \frac{r_{\mathrm{g}}^{3/2}}{2\sqrt{r_{\mathrm{s}}}}
     \ln \left| \frac{\sqrt{r} - \sqrt{r_{\mathrm{g}}}}{\sqrt{r} + \sqrt{r_{\mathrm{g}}}} \right|.
   \label{eq:genetortoisecoordinate}
 \end{align} 
We have chosen the integration constant so that $r_*=0$ at $r=0$ formally.\footnote{Since we are interested in the region~$r_{\rm g} < r < \infty$ and the function~$r_*(r)$ cannot be continuously extended beyond $r=r_{\rm g}$, it may be inappropriate to fix the integration constant at $r=0$ in a rigorous sense. 
Precisely speaking, we chose the integration constant so that
\begin{align*}
    r_{*} \to 2 \sqrt{\frac{r}{r_{\rm s}}} - \frac{\pi r_{\rm g}^{3/2}}{2\sqrt{r_{\rm s}}}
\end{align*}
as $r \to \infty$. 
In any case, the value of $r_*$ itself does not have a physical meaning, and we chose this particular value of integration constant just for simplicity (see Appendix~\ref{app:growingphase} for a related discussion).} 
Note that the generalized tortoise coordinate is regular at the Schwarzschild radius~$r=r_{\rm s}$. 
We also note that $r_{*} \to - \infty$ as $r \to r_{\mathrm{g}}$ and $r_{*} \to \infty$ as $r \to \infty$. 
This means that this coordinate plays the same role as the usual tortoise coordinate.
In the new coordinate system~$\{ \tilde {t}, r_{*}, \theta, \varphi\}$, the equation of motion for the monopole perturbations takes the form of a two-dimensional wave equation:
\begin{align}
  \frac{\partial^{2} \Psi}{\partial r_{*}^{2}} - \frac{\partial^{2} \Psi}{\partial \tilde t^{2}} - V_{\mathrm{eff}}(r_{*})\Psi = 0,
  \label{eq:Shcroeq}
\end{align}
where
\begin{align}
  V_{\mathrm{eff}} 
  = 
  \frac{\hat d_{3}}{\tilde d_{1}} + \frac{1}{(\tilde d_{1} \hat d_{2})^{1/4}} \frac{\mathrm{d}^{2}}{\mathrm{d}r_{*}^{2}} \brb{ (\tilde d_{1} \hat d_{2})^{1/4} }
  = 
  \frac{r_{\mathrm{s}}}{16r^{5}r_{\mathrm{g}}^{2}}
  (r^{2}-r_{\mathrm{g}}^{2})(3r^{2}-11r_{\mathrm{g}}^{2}),
\end{align}
is the effective potential. 
Figure~\ref{fig:Veffplot} shows the profile of the effective potential for $r_{\mathrm{g}}=0.5$ (blue dashed curve), $r_{\mathrm{g}}=1$ (orange dotted curve), and $r_{\mathrm{g}}=1.2$ (green solid curve). 

It is worth noting crucial differences of the effective potential from the standard Regge-Wheeler/Zerilli potential.
There are two roots for $V_{\mathrm{eff}}=0$ in the region~$r \geq r_{\mathrm{g}}$, which are given by $r_{\mathrm{g}}$ and $r_{0}=\sqrt{11/3}\, r_{\mathrm{g}}$. 
The effective potential approaches zero from above at infinity, while it reaches zero from below at the monopole horizon. 
This indicates that the effective potential takes negative values for $r_{\mathrm{g}}<r<r_{0}$, contrary to the standard Regge-Wheeler/Zerilli potential which remains positive outside the horizon. 
We also note that the effective potential well becomes deeper for smaller $r_{\rm g}$.

\subsection{\texorpdfstring{Property of a constant-$\tilde{t}$ hypersurface}{Character of a constant-tilde t hypersurface}}
\label{subsec:tildetconstsurface}

We have obtained the equation of motion for the monopole perturbations in the form of the wave equation by employing the new time coordinate~$\tilde{t}$ [see Eq.~\eqref{eq:newtimecoordinate}]. 
In solving the wave equation~\eqref{eq:Shcroeq}, we would like to impose initial conditions on a Cauchy surface. 
The well-posedness of the Cauchy problem is in general nontrivial and needs to be checked carefully, especially in modified gravity (see e.g., \cite{Motloch:2015gta}).
The Cauchy surface needs to satisfy several conditions (see, e.g., \cite{Wald:1984rg} and references therein).
In particular, it should be a spacelike hypersurface. 
Looking at the wave equation~\eqref{eq:Shcroeq}, one might be tempted to impose the initial condition on a $\tilde {t} = \mathrm{const.}$~surface. 
However, as we shall see below, 
$\tilde {t} = \mathrm{const.}$~surfaces are not necessarily spacelike. 

Let us consider a vector field~$n^{\mu}$ which is normal to a $\tilde {t} = \mathrm{const.}$ surface, i.e., $n_{\mu} \equiv \partial_{\mu} \tilde{t}$. 
The norm of $n^{\mu}$ is given by
\begin{align}
    n_{\mu} n^{\mu} = 
    \frac{(r^{5/2}-r_{\mathrm{g}}^{2}\sqrt{r_{\mathrm{s}}})(r^{3}+r_{\mathrm{g}}^{2}\sqrt{r_{\mathrm{s}}r})}{r_{\mathrm{s}}\sqrt{r}(r^{2}-r_{\mathrm{g}}^{2})^{2}}.
\end{align}
This means that $\tilde{t}=\mathrm{const.}$ surfaces are spacelike for $r<(r_{\mathrm{g}}^{4} r_{\mathrm{s}})^{1/5}$ and timelike for $r>(r_{\mathrm{g}}^{4} r_{\mathrm{s}})^{1/5}$. 
Now, we discuss the position of this characteristic radius~$(r_{\mathrm{g}}^{4} r_{\mathrm{s}})^{1/5}$ for {\it case~I}\,: $r_{\mathrm{g}}<r_{\mathrm{s}}$, 
{\it case~II}\,: $r_{\mathrm{g}}=r_{\mathrm{s}}$, and {\it case~III}\,: $r_{\mathrm{g}}>r_{\mathrm{s}}$, separately. 

\begin{itemize}
\item {\it case~I}\,: $r_{\mathrm{g}}<r_{\mathrm{s}}$\\
In this case, we have $r_{\mathrm{g}}<(r_{\mathrm{g}}^{4} r_{\mathrm{s}})^{1/5}<r_{\mathrm{s}}$. 
\begin{figure}[t]
\includegraphics[width=\textwidth]{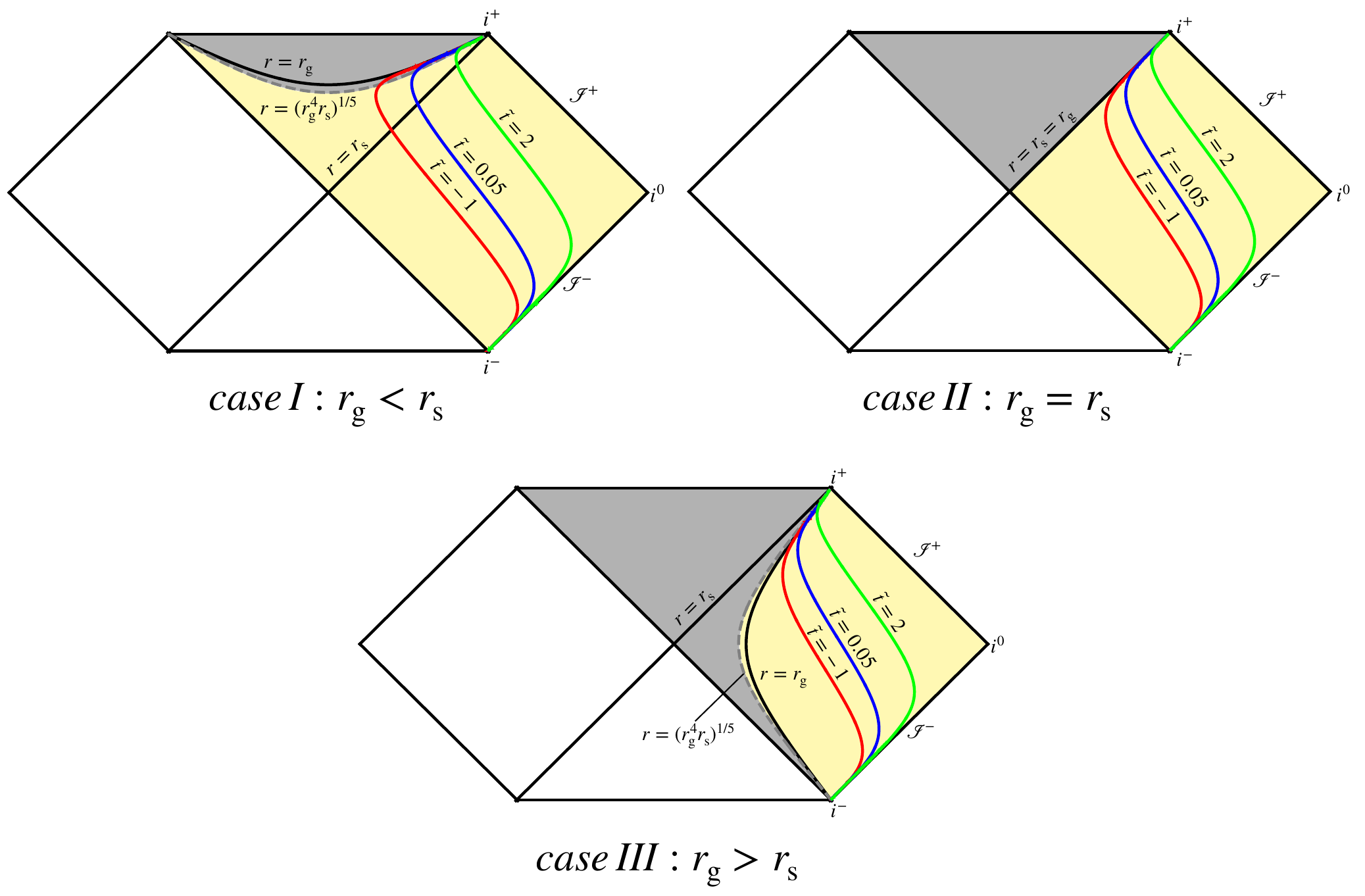}
\caption{Surfaces of constant~$\tilde{t}$ for the {\it cases~I, II}, and {\it III}. 
The red, blue, and green curves correspond to $\tilde{t}=-1$, $\tilde{t}=0.05$, and $\tilde{t}=2$, respectively. 
In the gray shaded region, $\tilde{t}=\mathrm{const.}$~surfaces are spacelike, while in the yellow shaded region, they are timelike. }
\label{fig:tildetconsts}
\end{figure}
Therefore, the region $r<(r_{\mathrm{g}}^{4} r_{\mathrm{s}})^{1/5}$ in which $\tilde{t}=\mathrm{const.}$~surfaces become spacelike lies inside the Schwarzschild radius. 
The left top panel in Fig.~\ref{fig:tildetconsts} shows $\tilde{t}=\mathrm{const.}$~surfaces for some specific values of $\tilde{t}$ for the {\it case I}. 
In the gray shaded region, $\tilde{t}=\mathrm{const.}$~surfaces are spacelike, while in the yellow shaded region, they are timelike. 
According to Fig.~\ref{fig:tildetconsts}, 
$\tilde{t}=\mathrm{const.}$~surfaces extend from the past timelike infinity~$i^{-}$ to the future timelike infinity~$i^{+}$. 
The important point is that $\tilde{t}=\mathrm{cosnt.}$~surfaces are not always spacelike.

\item {\it case~II} \,: $r_{\mathrm{g}}=r_{\mathrm{s}}$\\
In this case, we have $r_{\mathrm{g}}=(r_{\mathrm{g}}^{4} r_{\mathrm{s}})^{1/5}=r_{\mathrm{s}}$. 
As a result, 
$\tilde{t}=\mathrm{const.}$~surfaces are spacelike inside the horizon~$r=r_{\rm s}(=r_{\rm g})$, while they are timelike outside the horizon.
The top right panel in Fig.~\ref{fig:tildetconsts} shows $\tilde{t}= \mathrm{const.}$~surfaces in the {\it case II}. 
Since we focus only on the region of $r>r_{\mathrm{g}}$, $\tilde{t}=\mathrm{const.}$~surfaces are always timelike.

\item {\it case~III}\,: $r_{\mathrm{g}}>r_{\mathrm{s}}$\\
In this case, we have $r_{\mathrm{s}}<(r_{\mathrm{g}}^{4} r_{\mathrm{s}})^{1/5}<r_{\mathrm{g}}$.
The bottom panel in Fig.~\ref{fig:tildetconsts} shows $\tilde{t}=\mathrm{const.}$~surfaces in the {\it case III}. 
Unlike the {\it case I}, 
in the present case, the region of $r>r_{\mathrm{g}}$ is included in the yellow shaded region. 
Therefore, $\tilde{t}=\mathrm{const.}$~surfaces are always timelike. 
\end{itemize}

Therefore, $\tilde{t}=\mathrm{const.}$~surface is mostly timelike, and one cannot use it as the Cauchy surface.
We need to impose initial conditions on a truly spacelike hypersurface instead of a $\tilde{t}=\mathrm{const.}$~surface in order to consistently solve the initial value problems. 
We will discuss how to define a spacelike hypersurface in the present setup in detail in Sec.~\ref{sec:timeevolution}.

\subsection{Characteristic curves}
\label{subsec:characteristiccurves}

Before proceeding to the formulation of the initial value problem in Sec.~\ref{sec:timeevolution}, let us investigate the characteristic curves of the monopole perturbations about the stealth Schwarzschild solutions in the DHOST theory. 
The characteristic curves can be read off from the wave equation~\eqref{eq:Shcroeq}. 
In the high-frequency regime, 
the field~$\Psi$ propagates along the curves on which $\tilde{v} \equiv \tilde{t} + r_{*} = \mathrm{const.}$ or $\tilde{u} \equiv \tilde{t} - r_{*} = \mathrm{const.}$ is satisfied. 
From Eqs.~\eqref{eq:newtimecoordinate} and~\eqref{eq:genetortoisecoordinate}, 
we have 
\begin{align}
    \tilde{v} = 
    t 
    + 
    \frac{2}{3}\frac{r^{3/2}}{\sqrt{r_{\mathrm{s}}}}
    + 
    2 r_{\mathrm{g}} \sqrt{\frac{r}{r_{\mathrm{s}}}}
    +
    2 \sqrt{r_{\mathrm{s}}r}
    + 
    \frac{r_{\mathrm{g}}^{3/2}}{\sqrt{r_{\mathrm{s}}}} \ln \norm{ \frac{\sqrt{r} - \sqrt{r_{\mathrm{g}}}}{\sqrt{r} + \sqrt{r_{\mathrm{g}}}} }
    + 
    r_{\mathrm{s}} \ln 
    \left| \frac{\sqrt{r} - \sqrt{r_{\mathrm{s}}}}{\sqrt{r} + \sqrt{r_{\mathrm{s}}}} \right|,
    \label{eq:directionnomopolev}
    \\
    \tilde{u} =
    t 
    + 
    \frac{2}{3}\frac{r^{3/2}}{\sqrt{r_{\mathrm{s}}}}
    -
    2 r_{\mathrm{g}}\sqrt{\frac{r}{r_{\mathrm{s}}}}
    + 
    2 \sqrt{r_{\mathrm{s}} r}
    + \frac{2 r_{\mathrm{g}}^{3/2}}{\sqrt{r_{\mathrm{s}}}} \tan^{-1} \bra{ \sqrt{\frac{r}{r_{\mathrm{g}}}} }
    + 
    r_{\mathrm{s}} \ln 
    \left| \frac{\sqrt{r} - \sqrt{r_{\mathrm{s}}}}{\sqrt{r} + \sqrt{r_{\mathrm{s}}}} \right|.
    \label{eq:directionnomopoleu}
\end{align}
In terms of $\tilde{v}$ and $\tilde{u}$, the wave equation~\eqref{eq:Shcroeq} can be written as
\begin{equation}
    -4\frac{\partial^2\Psi}{\partial\tilde{u}\partial\tilde{v}}=V_{\rm eff}(r_{*})\Psi.
    \label{eq:Shcroeq_uv}
\end{equation}

\begin{figure}[t]
\includegraphics[width=\textwidth]{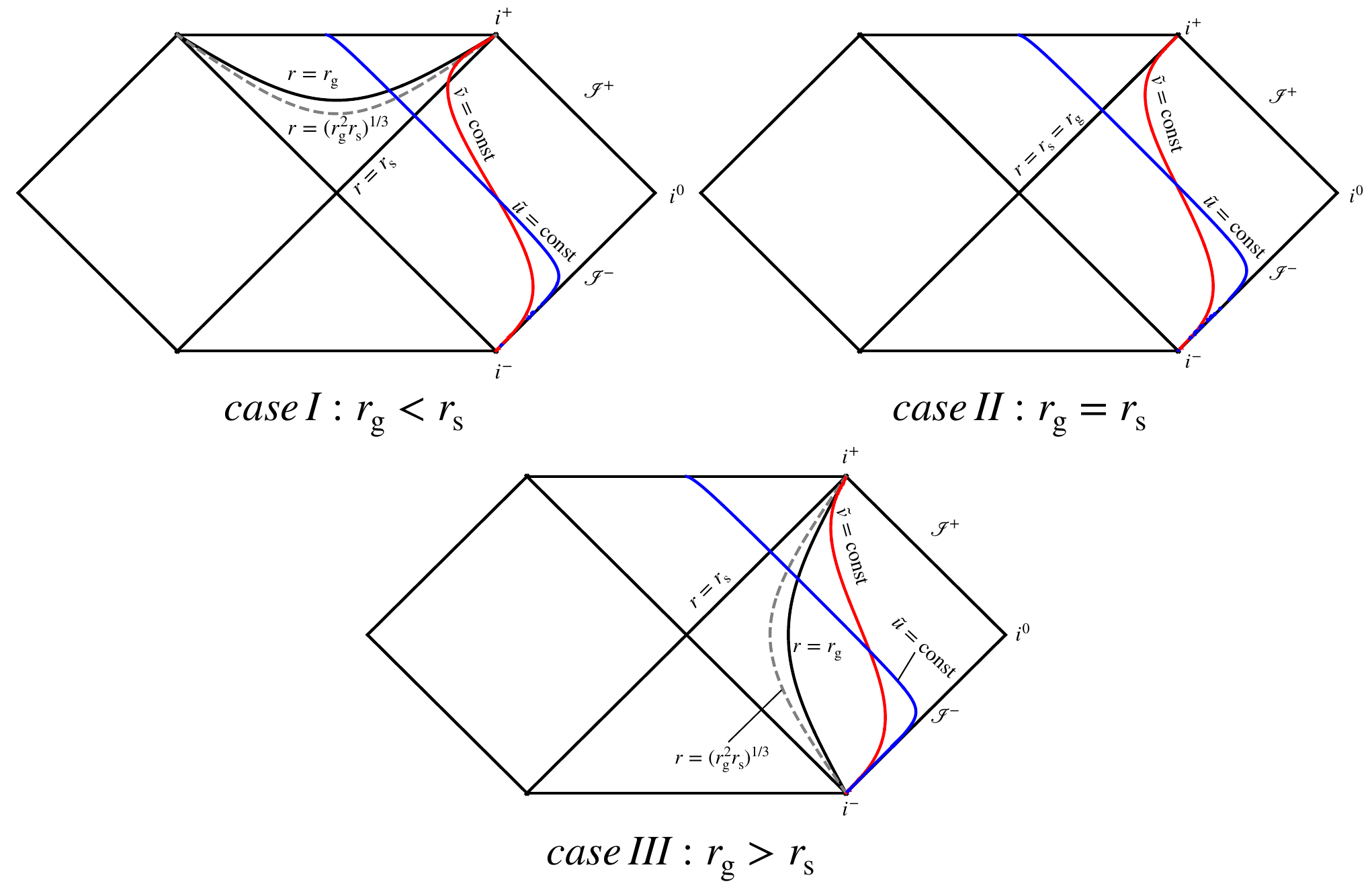}
\caption{Characteristic curves of the monopole perturbations for the {\it cases~I, II}, and {\it III}. 
The red and blue curves correspond to $\tilde{v}= \mathrm{const.} $ and $\tilde{u}= \mathrm{const.}$ curves, respectively. }
\label{fig:propagations}
\end{figure}

Let us investigate the property of the characteristic curves with respect to the background spacetime metric.
Performing an analysis similar to the one in the previous subsection, 
we can find that the characteristic curves become null at $r=(r_{\mathrm{g}}^{2}r_{\mathrm{s}})^{1/3}$. Figure~\ref{fig:propagations} shows typical plots of the characteristic curves of the monopole perturbations for the {\it cases~I, II}, and {\it III}. 
The $\tilde{u}=\mathrm{const.}$~curves extend from the past timelike infinity~$i^{-}$ to the spacetime singularity at $r=0$. 
On the other hand, 
the $\tilde{v}=\mathrm{const.}$~curves extend from the past timelike infinity~$i^{-}$ to the future timelike infinity~$i^{+}$. 
Thus, the monopole perturbations do not propagate along the null directions in the Schwarzschild spacetime and do not reach the future null infinity~$\mathscr{I}^{+}$.

We note that the monopole perturbations can be superluminal depending on the choice of parameters.
As mentioned earlier, the radius of the monopole horizon~$r_{\rm g}$ does not necessarily coincide with the Schwarzschild radius~$r_{\rm s}$.
For the {\it case I} with $r_{\rm g}<r_{\rm s}$, in the region of $r < (r_{\mathrm{g}}^{2}r_{\mathrm{s}})^{1/3}$, the monopole perturbations become superluminal, i.e., the characteristic curves become spacelike. 
However, as shown in top left panel in Fig.~\ref{fig:propagations}, characteristic curves cannot escape from the Schwarzschild radius once those enter the event horizon. 
Therefore, the monopole perturbations cannot carry information from the interior of the Schwarzschild radius. 
In Appendix~\ref{app:algorithmcurves}, we reanalyze the characteristics for the monopole perturbations in the DHOST theory by using an algorithm introduced in \cite{Motloch:2016msa}, which would be useful in a more general setup (e.g., scordatura DHOST theories). 
We confirm that the algorithm reproduces the result obtained in this subsection.

\section{Initial value problem and time evolution of the monopole perturbation}
\label{sec:timeevolution}

In this section, we shall formulate the well-posed initial value problem of the wave equation~\eqref{eq:Shcroeq_uv} for the monopole perturbations.
As mentioned above, in order to recast the equation of motion in the form of a two-dimensional wave equation, we needed to introduce a new time coordinate~$\tilde{t}$, which is different from the physical time. 
Therefore, quantities (e.g., quasinormal frequencies) computed by imposing a set of initial conditions on a $\tilde{t}=\mathrm{const}.$~surface may not be directly related to actual observables. 
Depending on the physical situation, we should carefully choose an appropriate initial surface. 
To have a better intuition, we discuss simple examples in Appendix~\ref{app:choiceinitialsurface}.

Another thing to note is that, in our analysis, we consider an idealized situation where the matter sector is ignored.
However, from a physical point of view, one should take into account other fields as matter fields in reality.
Provided that the matter fields are minimally coupled to gravity, a field propagating with the speed of light travels along a 45-degree path on the Penrose diagram, which defines the causal structure of the spacetime.
Since we impose initial conditions for the matter fields on a spacelike hypersurface with respect to the background metric, we should also impose initial conditions for the metric (and the scalar field if we employ a non-unitary gauge) on a spacelike hypersurface.
Otherwise, one would impose initial conditions for the monopole perturbations on a timelike hypersurface and evolve those towards a spacelike direction, which implies that the evolution of the monopole perturbations violate the causality defined by a field propagating with speed of light.
Also, the Cauchy problem would be ill-posed in the sense that one cannot give initial conditions for all the fields in a single spacelike hypersurface.\footnote{Note that we encounter a similar situation when we consider the Ostrogradsky ghost in higher-derivative theories~\cite{Woodard:2015zca}.
Without the interaction to other fields, a system with negative energy is totally no problem and there is no instability.
However, we implicitly assume the interaction between the ghosty gravitational sector and other matter fields which possess the positive energy, leading to the instabilities.
Therefore, we avoid ghost degree(s) of freedom and work in degenerate theories.
From a similar point of view, here we avoid to use a constant-$\tilde t$~surface, that is not spacelike, as the initial surface.}

Having clarified these points, we define a physically sensible formulation of the initial value problem by explicitly constructing a spacelike hypersurface on which we impose the initial conditions.
We also apply our formulation to the time evolution of a Gaussian wave packet, along the same lines as~\cite{Vishveshwara:1970zz}.

\subsection{Formulation of well-posed initial value problem}

As we saw in Sec.~\ref{subsec:tildetconstsurface},
a hypersurface of constant $\tilde{t}=(\tilde{u}+\tilde{v})/2$ is not spacelike, and hence it is not appropriate as an initial surface.
Therefore, we introduce new coordinates~$\tilde{U} \equiv a \tilde{u}$ and $\tilde{W} \equiv - b \tilde{v}$,\footnote{The sign flip in the definition of $\tilde{W}$ is necessary so that the coordinate value increases as the field evolves in the numerical calculation.} where $a$ and $b$ are positive constants chosen so that a hypersurface of constant $\tilde{U}+\tilde{W}$ (which we call $\tilde{\Sigma}$) is spacelike at least in some finite region (say $S$).\footnote{The hypersurface~$\tilde{\Sigma}$ depends only on the ratio~$a/b$, and one can fix either $a$ or $b$ without loss of generality. However, we keep both the parameters since this redundancy is technically convenient for optimizing the size of the spacelike region.}
Strictly speaking, the hypersurface~$\tilde{\Sigma}$ itself is not an initial surface we would like to construct, but it will be used below to design an initial surface which is spacelike everywhere.
In terms of the new coordinates, the wave equation~\eqref{eq:Shcroeq_uv} can be written as
\begin{align}
    -4 \frac{\partial^{2} \Psi}{\partial \tilde{U} \partial \tilde{W}} = \tilde{V}(r_{*}) \Psi.
    \label{eq:doublenullequation}
\end{align} 
Here, we have defined the new potential~$\tilde{V}$ by  
\begin{align}
    \tilde{V} 
    \equiv - \frac{V_{\mathrm{eff}}}{ab}
    = 
    -\frac{r_{\mathrm{s}}}{16 ab r_{\mathrm{g}}^{2} r^{5}}
  (r^{2}-r_{\mathrm{g}}^{2})(3r^{2}-11r_{\mathrm{g}}^{2}), \label{eq:Vtilde}
\end{align}
where $r$ is a function of the generalized tortoise coordinate $r_{*}$ whose expression is given by Eq.~\eqref{eq:genetortoisecoordinate}. 
\begin{figure}[t]
\includegraphics[width=120mm]{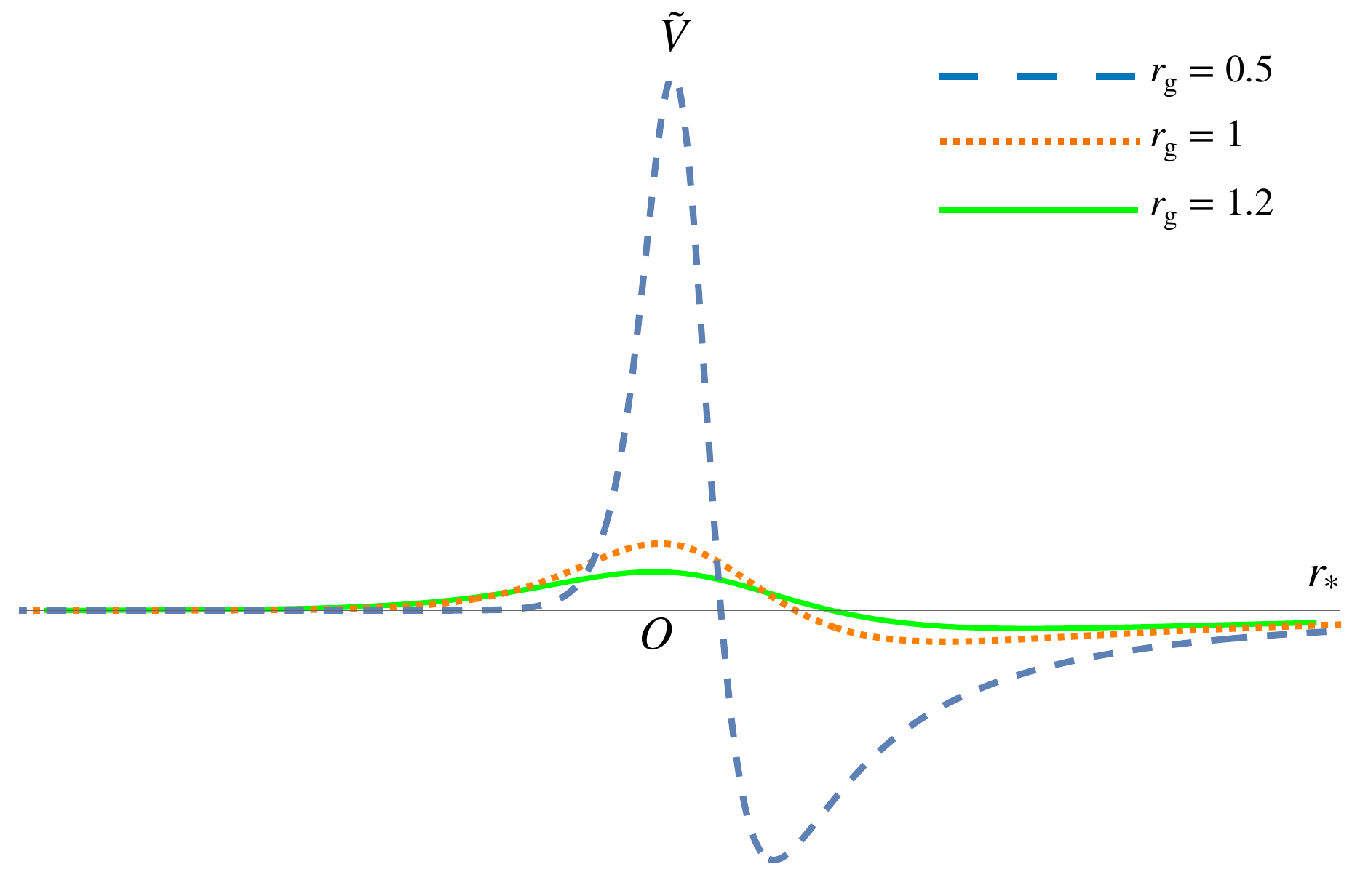}
\caption{Typical plots of the potential~$\tilde{V}(r_{*})$ for $r_{\mathrm{g}}=0.5$ (blue dashed curve), $r_{\mathrm{g}}=1$ (orange dotted curve), and $r_{\mathrm{g}}=1.2$ (green solid curve) in units of $r_{\rm s}$.}
\label{fig:potentialrstar}
\end{figure}
Figure~\ref{fig:potentialrstar} shows typical plots of the potential~$\tilde{V}$ as a function of $r_{*}$. 
Due to the sign change from $V_{\mathrm{eff}}$, $\tilde{V}$ is negative for large $r_{*}$.
Therefore, when we consider the scattering of a wave packet coming from the spatial infinity, the wave packet passes the negative potential region first, and then is scattered by the positive potential barrier.

\begin{figure}[t]
\includegraphics[width=\textwidth]{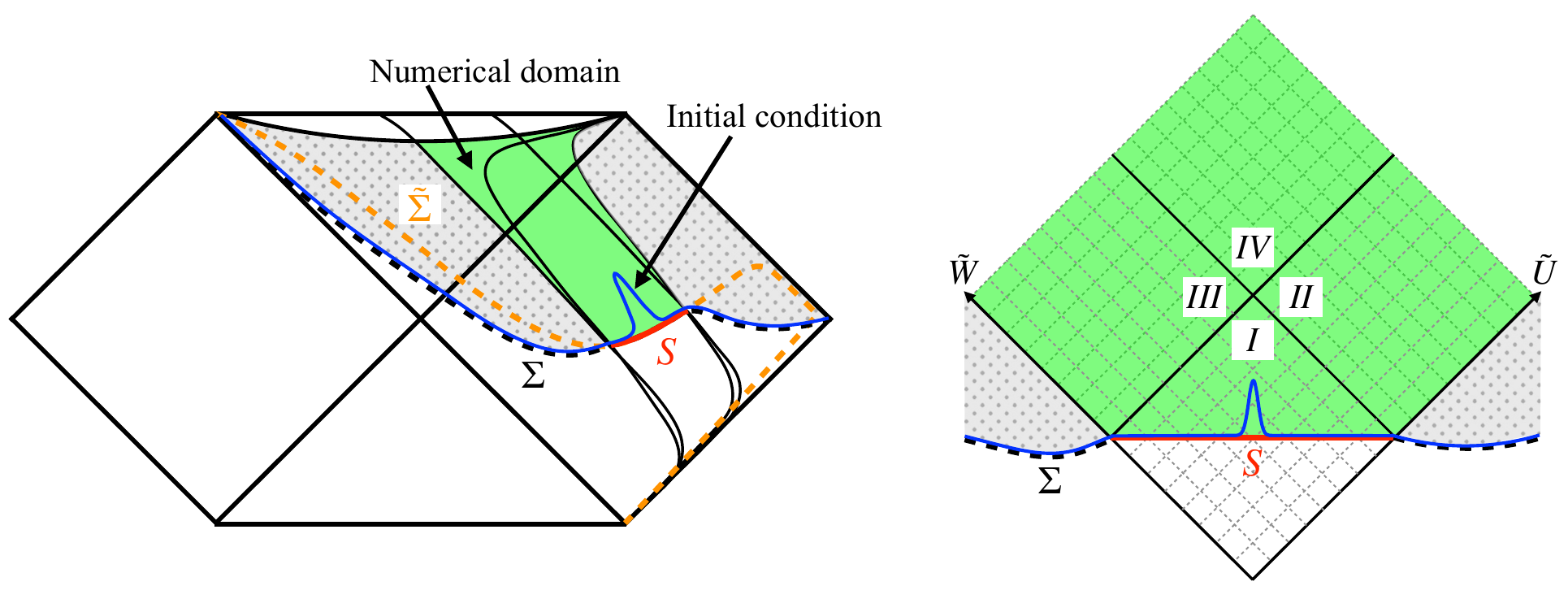}
\caption{Schematic picture of a initial surface~$\Sigma$ (black dashed curve) and $\tilde{U}+\tilde{W}=\mathrm{const.}$~surface~$\tilde{\Sigma}$ (orange dashed curve). 
We impose initial conditions (blue solid curve) on the spacelike hypersurface~$\Sigma$. 
We require that the surface~$\Sigma$ coincides with $\tilde{\Sigma}$ in the region~$S$ (red solid curve). 
We also require that the initial field values $\Psi_{\rm ini}$ have a compact support in $S$, i.e., they are nonvanishing only inside the numerical domain (green shaded region). 
Outside the numerical domain (gray shaded region), 
the field vanishes because we impose that the derivative in the direction perpendicular to $\Sigma$ is zero on $\Sigma$. 
In the region $S$, this condition corresponds to $(\partial_{\tilde{U}}+\partial_{\tilde{W}})\Psi|_{S}=0$.}
\label{fig:initialsurface}
\end{figure}
Figure~\ref{fig:initialsurface} shows a schematic picture of the initial surface and the numerical domain. 
The left panel shows our numerical setup in the Penrose diagram, while the right panel shows it in a diagram where the characteristic curves of the monopole perturbations are 45-degree straight lines. 
Let $\Sigma$ denote a spacelike hypersurface on which we impose initial conditions.
We impose the following two requirements on the hypersurface~$\Sigma$ and the initial conditions:
\begin{enumerate}
\renewcommand{\theenumi}{\roman{enumi}}
\renewcommand{\labelenumi}{(\theenumi)}
\item \label{req1} The initial surface~$\Sigma$ coincides with $\tilde{\Sigma}$ in a region~$S$ within the numerical domain.
\item \label{req2} The initial conditions have a compact support in $S$.
\end{enumerate}
Since we focus on the causal future of the region~$S$ determined by the characteristic curves of the monopole perturbations, under the requirement~(\ref{req1}), imposing an initial condition on $\tilde{\Sigma}$ corresponds to imposing an initial condition on $\Sigma$. 
Under this requirement, we can regard $\tilde{U}+\tilde{W}$ as the physical time since a hypersurface of constant $\tilde{U}+\tilde{W}$ is spacelike in the region~$S$. 
Meanwhile, $\tilde{U}-\tilde{W}$ does not correspond to the generalized tortoise coordinate~$r_*$ since $r_*=-(\tilde{U}/a+\tilde{W}/b)/2$ with $a\ne b$. 
Therefore, the potential~$\tilde{V}(r_*)$ depends on both $\tilde{U}-\tilde{W}$ and $\tilde{U}+\tilde{W}$, and hence it can be regarded as ``time''-dependent, which is another crucial difference from the initial value problem for the standard Regge-Wheeler/Zerilli potential.
Then, the initial data in the region~$S$ determine the evolution in the region~I in the right panel of Fig.~\ref{fig:initialsurface}. 
The requirement~(\ref{req2}) is necessary to extend the numerical domain. 
Thus, as the initial data, we consider a Gaussian wave packet with a truncation within the region~$S$ to make its support compact. 
We also assume that the derivative in the direction perpendicular to $\Sigma$ is zero on the initial surface~$\Sigma$. 
In the region~$S$, this means that $(\partial_{\tilde{U}}+\partial_{\tilde{W}})\Psi|_{S}=0$. 
With the requirement~(\ref{req2}), in the gray shaded region in Fig.~\ref{fig:initialsurface}, the field vanishes. 
Consequently, $\Psi=0$ on both the right boundary of the region~II and the left boundary of the region~III. 
Combining these field values on the boundaries of the regions~II and III with the solution in the region~I, we can extend the numerical domain to the regions~II, III, and IV.

\subsection{Time evolution of a Gaussian wave packet}
\label{subsec:scattering}

In this subsection, we examine the time evolution of the monopole perturbations based on the wave equation~\eqref{eq:doublenullequation}. 
We employ the numerical method introduced in~\cite{Gundlach:1993tp,Lucietti:2012xr}. 
We discretize both Eq.~\eqref{eq:doublenullequation} and the condition~$(\partial_{\tilde{U}}+\partial_{\tilde{W}})\Psi|_{S}=0$ with the coordinates~$\tilde{U}$ and $\tilde{W}$. 
As we discuss in Appendix~\ref{app:method}, we confirm that the numerical method we use shows ${\cal O}(h^{2})$ convergence, where $h$ is the grid size.

In order to discuss the dynamical stability of the stealth Schwarzschild solutions in the DHOST theory, we analyze the Vishveshwara's classical scattering experiment~\cite{Vishveshwara:1970zz}, which is the time evolution of a Gaussian wave packet. 
Here, we focus on the {\it case~I} where $r_{\rm g}<r_{\rm s}$ and exemplify the results with $r_{\mathrm{g}} = 0.5 \,r_{\mathrm{s}}$. 
For the {\it cases~II} and {\it III}, we show typical results in Appendix~\ref{app:resultsother}, which are actually qualitatively similar to the one for the {\it case I}. 
Hereafter, we set $r_{\mathrm{s}}=1$ for simplicity. 
We set the scaling constants as $a=5$, $b=4.8$, and the numerical domain as
$-175 \le \tilde{U} \le 105$ and $-70 \le \tilde{W} \le 210$. 
We choose the grid size as $h=0.04$. 
\begin{figure}[t]
\includegraphics[width=150mm]{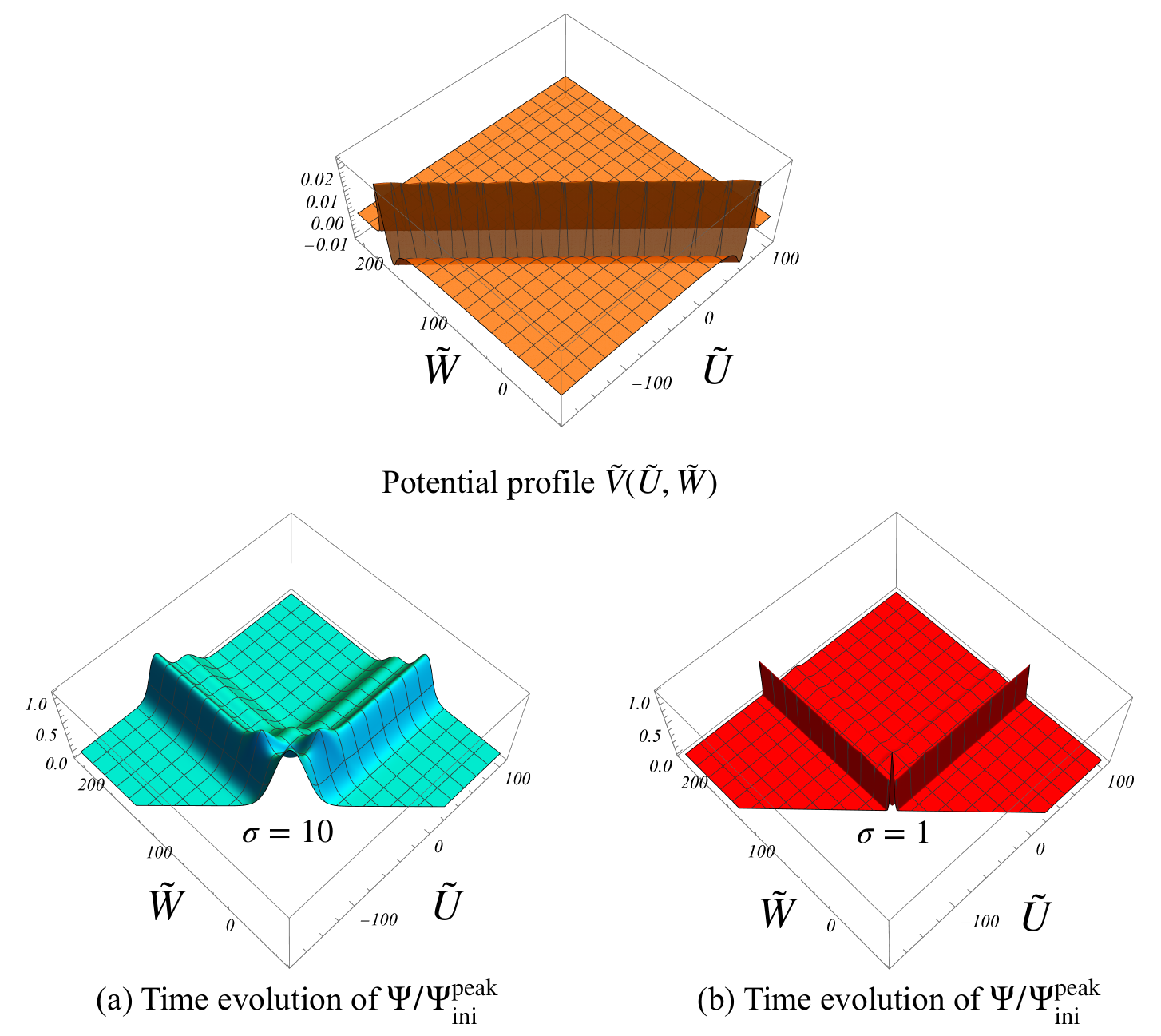}
\caption{
The potential profile in the $(\tilde{U},\tilde{W})$-space (top panel) and the time evolution of the monopole perturbations for the case~$r_{\mathrm{g}}=0.5$ in the $(\tilde{U},\tilde{W})$-space with (a)~$\sigma=10$ and (b)~$\sigma=1$.
}
\label{fig:timeevolution05}
\end{figure}

\begin{figure}[t]
\includegraphics[width=\textwidth]{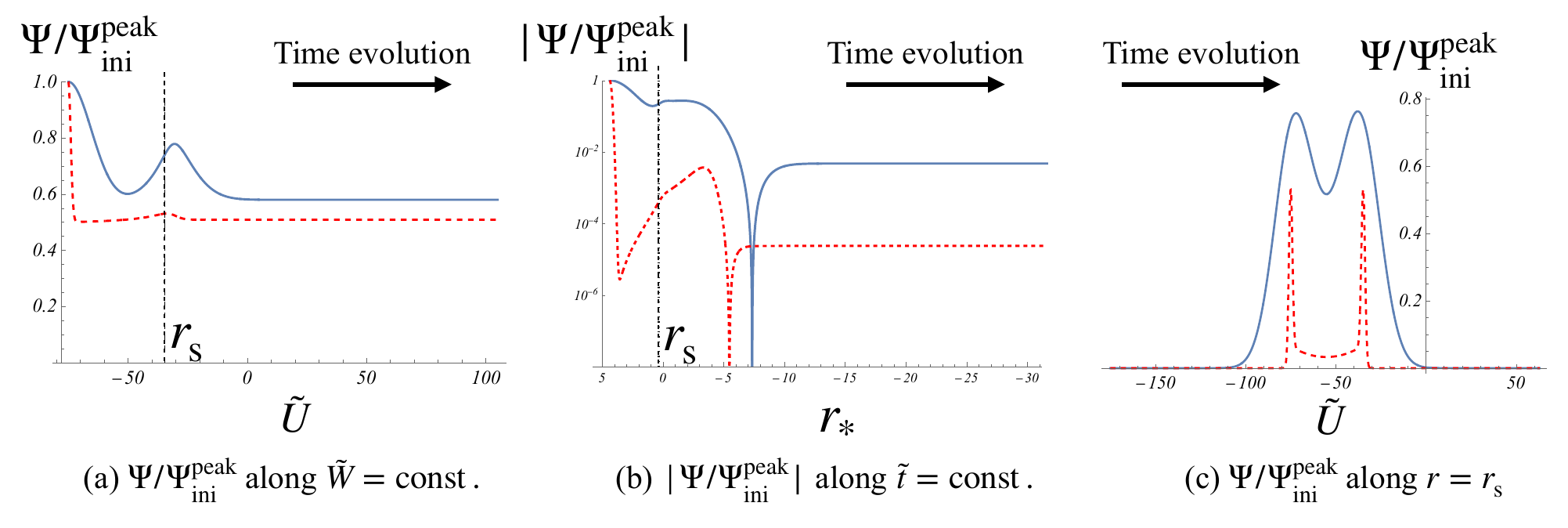}
\caption{
Time evolution of the monopole perturbations for $r_{\mathrm{g}}=0.5$ along (a) $\tilde{W}=\mathrm{const}.$, (b) $\tilde{t}=\mathrm{const}.$, and (c) $r=r_{\mathrm{s}}$, respectively.
The blue solid curve and the red dashed curve respectively depict the time evolution for $\sigma=10$ and $\sigma=1$. }
\label{fig:timeevolution05const}
\end{figure}

In Fig.~\ref{fig:timeevolution05}, we show 
the potential profile in the $(\tilde{U},\tilde{W})$-space (top panel), and 
the time evolution of the monopole perturbations in the $(\tilde{U},\tilde{W})$-space with different widths of the initial Gaussian wave packets: (a) $\sigma = 10$ and (b) $\sigma =1$. 
In these figures, we show the values of the monopole perturbations normalized by $\Psi_{\rm ini}^{\rm peak}$ defined in Eq.~\eqref{eq:psipeak}. 
On the other hand, Figs.~\ref{fig:timeevolution05const}(a) and \ref{fig:timeevolution05const}(b) show the time evolution\footnote{Let us clarify the terminology of ``time evolution.'' In Fig.~\ref{fig:timeevolution05const}, we depict the direction of time evolution. 
Along this direction, the coordinate value of $t$, which is the Killing time, increases for $r>r_{\mathrm{s}}$, while the coordinate value of $r$, which is the areal radius, decreases for $r<r_{\mathrm{s}}$. 
For $r=r_{\mathrm{s}}$, the coordinate value of the ingoing Eddington-Finkelstein coordinate $v$ increases along the time evolution. 
Therefore, the direction of ``time evolution'' has the usual meaning even for the present case.} of the monopole perturbations along $\tilde{W}=\mathrm{const.}$ and $\tilde{t}=\mathrm{const.}$, respectively. 
The left ends of the curves in Figs.~\ref{fig:timeevolution05const}(a) and \ref{fig:timeevolution05const}(b) match the peak of the initial data~$\Psi_{\rm ini}^{\rm peak}$. 
Figure~\ref{fig:timeevolution05const}(c) shows the evolution along $r=r_{\mathrm{s}}$. In Figs.~\ref{fig:timeevolution05const}(a)--\ref{fig:timeevolution05const}(c), we show the values or the absolute values of the monopole perturbations normalized by $\Psi_{\rm ini}^{\rm peak}$.

Firstly, we explain the evolution along the characteristic curves.
From Figs.~\ref{fig:timeevolution05}(a) and~\ref{fig:timeevolution05}(b), we can see that the initial Gaussian wave packet splits into two directions, which are along the characteristic curves: $\tilde{U}=\mathrm{const.}$ and $\tilde{W}=\mathrm{const.}$ directions. 
This behavior is nothing but the one for the d'Alembert solution for the wave equation.
Then, evolving the monopole perturbations along the characteristic curves, a growing phase appears [see Fig.~\ref{fig:timeevolution05const}(a)]. 
The reason for the appearance of the growing phase is that the monopole perturbations initially propagate in the negative-potential region, where the dependence of the potential on $r_{*}$ is given by $\tilde{V} \sim -r_{*}^{-2}$ as $r_{*} \to \infty$. 
We analyze the growing phase in detail in Appendix~\ref{app:growingphase}. 
It is worth noting that the analysis in Appendix~\ref{app:growingphase} shows that the growth rate of the monopole perturbations is power-law. 
Moreover, this growing phase does not continue for a long time because the monopole perturbations will enter the positive-potential region. 
After entering the positive-potential region, the potential decays exponentially and the system rapidly approaches the massless Klein-Gordon system. 
Consequently, in the positive-potential region, the monopole perturbations do not decay completely and continue to propagate along the characteristic curves even if we extend the numerical domain. One wave packet propagates towards $i^{+}$, and the other towards the spacetime singularity.

Secondly, let us discuss the evolution along $\tilde{t}=\mathrm{const.}$ [see Fig.~\ref{fig:timeevolution05const}(b)]. 
We note that $b \tilde{U}-a \tilde{W}=\mathrm{const.}$ along a curve with $\tilde{t}=\mathrm{const}$.
According to Fig.~\ref{fig:timeevolution05const}(b), the monopole perturbations decay initially, but a growing phase follows.
The origin of this growing phase is that the waves propagating along the characteristic curves are scattered by the potential. 
The scattered waves change their directions and propagate again along the characteristic curves. 
The growing phase appears as the central bump in Figs.~\ref{fig:timeevolution05}(a) and~\ref{fig:timeevolution05}(b). 
After the growing phase, the monopole perturbations decay. 
Eventually, the field value becomes a negative constant. 
This asymptotic behavior is consistent because the system is almost the massless Klein-Gordon system, which allows a constant solution as a trivial solution.
Note that the asymptotic field value at the late time is the same for both $\sigma=10$ and $\sigma=1$ cases, which cannot be seen in Fig.~\ref{fig:timeevolution05const}(b) though, because the two plots are normalized by different values of $\Psi_{\rm ini}^{\rm peak}$. Specifically, the asymptotic field value is about $-4\times 10^{-3}$.
Actually, the asymptotic field value is related to the area of the initial Gaussian wave packet. 
The reason why the cases with different values of $\sigma$ share the same value of the asymptotic field value is that we normalized the initial Gaussian wave packet by its area.

Finally, we explain the evolution along $r=r_{\mathrm{s}}$ shown in Fig.~\ref{fig:timeevolution05const}(c). 
As time goes by, the field experiences two peaks. 
These two peaks correspond to the wave packets propagating along the characteristic curves in the negative-potential region. 
Therefore, the height of the two peaks in Fig.~\ref{fig:timeevolution05const}(c) match the field value in Fig.~\ref{fig:timeevolution05const}(a) at $r=r_{\mathrm{s}}$, which can be read off from the intersections between the black dashed line and the blue/red curves in Fig.~\ref{fig:timeevolution05const}(a). 
According to Fig.~\ref{fig:timeevolution05const}(c), the amplitude of the monopole perturbations is finite and does not diverge along the event horizon.

We make a remark on the stability of the stealth Schwarzschild solutions in the DHOST theory against the monopole perturbations. 
Our numerical calculations confirmed that the amplitude of the monopole perturbations does not grow exponentially. The potential~$\tilde{V}$ has the negative region far outside the black hole, where a growing phase of the monopole perturbations appears. 
However, the growing phase does not continue for a long time because the monopole perturbations enter the positive-potential region. 
We find that, along the characteristic curves, the monopole perturbations do not decay completely and one wave packet keeps propagating towards the future timelike infinity~$i^{+}$ and the other towards the spacetime singularity. 
Naively, this result may seem to imply that the amplitude near $i^{+}$ increases as the monopole perturbations gather. 
However, this is not the case because the characteristic curves do not cross each other until they reach $i^{+}$.
On the other hand, the amplitudes of long-wavelength modes could become large, 
as we discuss in Appendix~\ref{app:growingphase}. 
However, in our numerical calculations, we cannot arbitrarily extend the region~$S$ where we give the initial conditions because the range in which the surface~$\tilde{\Sigma}$ remains spacelike is not so wide. 
Consequently, it would be challenging to perform numerical calculations for an initial Gaussian wave packet with large width. 
There remains a possibility that the stealth Schwarzschild solutions in the DHOST theory may be unstable for the monopole perturbations with long wavelength,  
although such long wavelength instabilities, even if they exist, would be harmless, similar to the Jeans instability.

Here, we discuss another simple choice of the initial surface. 
A constant-$\tau$ surface, where $\tau$ is the time coordinate in the Lema\^itre coordinate system, is spacelike as we mentioned in Sec.~\ref{subsec:stealthsolution}. 
Therefore, a constant-$\tau$ surface is also appropriate for the initial surface of the Cauchy problem. 
However, we did not choose a constant-$\tau$ surface as the initial surface because the coordinate~$\tau$ cannot be expressed in terms of $\tilde{U}$ and $\tilde{W}$ in a simple form.
In solving the wave equation~\eqref{eq:doublenullequation}, an initial surface defined by $\tilde{U}+\tilde{W}=\mathrm{const.}$ is more convenient for our numerical calculation.
Nevertheless, even if we choose a constant-$\tau$ surface as the initial surface, we expect that the qualitative behaviors of the time evolution of the monopole perturbations do not change as long as we consider the initial condition with a compact support.

Let us summarize physical interpretation of our results.
Our numerical calculations confirmed that the amplitude of the monopole perturbations does not grow exponentially. 
It implies that the stealth Schwarzschild solutions in the DHOST theory is dynamically stable against the monopole perturbations with the wavelength comparable or shorter than the size of the black hole horizon. 
Another important point is that the damped oscillations at the late time do not show up in our calculation unlike the ringdown phase in the standard case of general relativity, and the field value asymptotically approaches a constant. 
As we mentioned above, this asymptotic behavior is consistent because the system is almost the massless Klein-Gordon system in the asymptotic region.

\subsection{Static perturbation}
\label{subsec:staticmonopole}

While we focused on the dynamical behaviors of the perturbations so far,
in this subsection, we briefly discuss the static monopole perturbations
which may also contain physically important solutions.
The detailed calculation is shown in Appendix~\ref{app:staticmode}.
Imposing the regularity condition at $r = r_{\mathrm{g}}$, the only allowed solution
corresponds to the shift of the mass parameter of the Schwarzschild spacetime.
On the other hand,
if we do not impose the regularity condition at $r = r_{\mathrm{g}}$,
the general solution of the static perturbation contains monopole scalar hair, and the perturbed metric is singular at $r = r_{\mathrm{g}}$.
As far as we checked in our numerical calculation of the time evolution of the monopole perturbations, the singular behavior near $r=r_{\mathrm{g}}$ does not appear. 
This suggests that the static spherically symmetric spacetime with the nontrivial monopole scalar hair is not realized
in a dynamical formation process.

\section{Summary and discussions}
\label{sec:summary}

In the present paper, we studied perturbations about stealth Schwarzschild solutions with a linearly time-dependent scalar field in shift- and reflection-symmetric DHOST theories.
In particular, we focused on the monopole perturbation for simplicity and derived the master equation in the form of a two-dimensional wave equation to investigate its time evolution [see Eq.~\eqref{eq:Shcroeq}].
In doing so, we introduced a new time coordinate~$\tilde{t}(t,r)$ as in Eq.~\eqref{eq:newtimecoordinate}, with $t$ and $r$ being the Schwarzschild Killing time and radial coordinates.
A remarkable fact is that a hypersurface of constant~$\tilde{t}$ is not spacelike and hence is not appropriate as an initial surface of the Cauchy problem.
Therefore, in order to formulate the initial value problem in a physically sensible manner, we defined a spacelike hypersurface~$\Sigma$ on which we impose initial conditions.
More concretely, we first constructed a hypersurface~$\tilde{\Sigma}$ by slightly tilting a $\tilde{t}=\mathrm{const}.$~surface such that $\tilde{\Sigma}$ is spacelike at least in some finite region (say $S$) and required that \eqref{req1}~the initial surface~$\Sigma$ coincides with $\tilde{\Sigma}$ in the region~$S$ and \eqref{req2}~the initial conditions have a compact support in the region $S$.
By using our formulation, we analyzed the time evolution of a Gaussian wave packet, which is known as the Vishveshwara's classical scattering experiment.
As a result, we confirmed that stealth Schwarzschild solutions in the DHOST theory are dynamically stable against the monopole perturbations with the wavelength comparable or shorter than the size of the black hole horizon. 
We also found that the damped oscillations at the late time do not show up unlike the ringdown phase in the standard case of general relativity.

A general lesson is that, to formulate the well-posed Cauchy problem, it is necessary to elucidate the nature of the coordinates in the global structure of the spacetime. 
One cannot discern a priori an appropriate time coordinate to define the spacelike initial surface by just looking at the form of the equation of motion.

Although we demonstrated our idea in the case of the monopole perturbation about the stealth Schwarzschild solutions in DHOST theories in the present paper, the formulation has wide applications.
It would be applicable to, e.g., higher multipoles~$\ell\ge 2$ for which additional modes corresponding to gravitational waves are present.
Moreover, our formulation would apply to a broader class of solutions and/or theories.
Specifically, it would be intriguing to incorporate the scordatura term in the Lagrangian, by which the strong coupling problem for stealth solutions in DHOST theories can be circumvented~\cite{Motohashi:2019ymr,DeFelice:2022xvq}.
Another important thing to investigate is the connection to observables.
For instance, in general relativity, it is known that the ringdown phase of black hole merger is well described by the superposition of the quasinormal modes.
However, it remains unknown whether or not the same holds for modified gravity theories in general.
Therefore, besides the quasinormal modes, one should also analyze the time evolution of perturbations to clarify whether or not the quasinormal modes are excited and they dominate the late-time behavior of the perturbations, in which our formulation would play a key role.
We leave these issues for future work.

\acknowledgments{
We would like to thank T.~Harada, T.~Ishii, T.~Kobayashi, Y.~Mishima, Y.~Nakayama, N.~Sago, N.~Tanahashi, and K.~Tomikawa for useful comments and discussions. 
This work was supported by JSPS KAKENHI Grant Nos.~JP20H04746~(M.K.), JP22K03626~(M.K.), JP18K13565~(H.M.), JP22K03639~(H.M.), and JP21J00695~(K.T.) from the Japan Society for the Promotion of Science.
}

\appendix

\section{Strong coupling and the scordatura effect}
\label{app:scordatura}

In Sec.~\ref{ssec:monopole_master}, we showed that the sound speed of the monopole perturbation about the stealth Schwarzschild solution vanishes in the asymptotic flat region [see Eq.~\eqref{eq:soundspeedDHOST}].
Interestingly, a similar situation occurs for stealth black holes in a wide class of DHOST theories~\cite{Babichev:2018uiw,Minamitsuji:2018vuw,deRham:2019gha,Takahashi:2021bml}.
As we see below, the vanishing sound speed signals strong coupling and this issue can be circumvented by taking into account the so-called scordatura effect~\cite{Motohashi:2019ymr}.\footnote{Further, the scordatura is also necessary to make the quasi-static limit of the matter density perturbation well-defined, which implies that the subhorizon observables are inevitably affected
by the scordatura~\cite{Gorji:2020bfl}. 
The scordatura term also plays an important role in inflationary scenarios with small sound speed in the context of the production of primordial black holes~\cite{Gorji:2021isn}. }

The origin of the strong coupling issue (as well as a way out) can be explained in an EFT language.
In the context of ghost condensation, the authors of \cite{ArkaniHamed:2003uy} constructed an EFT of perturbations about the stealth Minkowski spacetime where the time diffeomorphism invariance is spontaneously broken by a spatially uniform scalar field.
Given the symmetry breaking pattern, it is straightforward to find the effective action of the associated Nambu-Goldstone boson.
Taking the decoupling limit of the effective action, the dispersion relation for the Nambu-Goldstone mode can be read off as
\begin{align}
    \frac{\omega^2}{\Lambda^2}=s_4\frac{k^4}{\Lambda^4}+s_6\frac{k^6}{\Lambda^6}+\cdots, \label{disp_EFT}
\end{align}
where $\Lambda\,(\ll \MPl)$ is a cutoff scale and the coefficients~$s_4$, $s_6$, $\cdots$ are dimensionless constants of order unity.
Note that the $k^2$~term is absent due to the symmetry of the background spacetime and the scalar field profile. 
It should also be noted that the terms of higher order in $k$, which correspond to higher spatial derivatives, never show up if the stealth Minkowski solution is realized in DHOST theories where the equations of motion are intrinsically of second order.
This means that the sound speed vanishes, which is nothing but what we found in the asymptotic flat region of the stealth Schwarzschild solution in DHOST theories.
The vanishing sound speed signals strong coupling because the strong coupling scale is proportional to some positive power of the sound speed~\cite{Cheung:2007st}.
Therefore, the strong coupling issue for stealth solutions is universal in DHOST theories.
In other words, in order to avoid the strong coupling issue within the framework of scalar-tensor theories, we need to take into account higher-derivative terms that violate the degeneracy.\footnote{Another possibility is to consider theories without propagating scalar degree of freedom (e.g., the cuscuton model~\cite{Afshordi:2006ad} and its extension~\cite{Iyonaga:2018vnu,Iyonaga:2020bmm}), for which the strong coupling problem is intrinsically absent.
Note also that there is no strong coupling problem for stealth solutions in degenerate vector-tensor theories~\cite{Mukohyama:2006mm,Aoki:2021wew}.}
Of course, the detuning from DHOST theories revives Ostrogradsky ghost in general, but the ghost is harmless so long as its mass scale is higher than the cutoff.
In fact, from the EFT point of view, the degeneracy may well be violated since it is not protected by symmetry in general.
Such a controlled detuning of the degeneracy was dubbed the scordatura effect in \cite{Motohashi:2019ymr}.

We briefly review how the scordatura effect works, following the logic of \cite{Motohashi:2019ymr,DeFelice:2022xvq}.
As an illustrative example, let us consider the following Lagrangian:
\begin{align}
    {\cal L} = {\cal L}_{\mathrm{DHOST}} + \alpha \frac{(\Box \phi)^{2}}{\Lambda^{2}}, \label{scordatura}
\end{align}
where $\Lambda\,(\ll\MPl)$ is a mass scale and $\alpha$ is a dimensionless constant of order unity.
Here, ${\cal L}_{\mathrm{DHOST}}$ denotes the Lagrangian of a DHOST theory in \eqref{eq:actionDHOST} satisfying the degeneracy conditions~\eqref{eq:degeneracycondition}, and the second term multiplied by $\alpha$ violates the degeneracy.
In other words, the parameter~$\alpha$ characterizes the deviation from DHOST theories (and controls the scordatura effect as we shall see below).
The Ostrogradsky ghost shows up above the scale~$\Lambda$, and hence the theory described by the Lagrangian~${\cal L}$ should be regarded as an EFT valid up to the scale~$\Lambda$.
This theory admits the Minkowski spacetime with a linearly time-dependent scalar field,
    \begin{equation}
    \bar{g}_{\mu\nu} \mathrm{d} x^\mu \mathrm{d} x^\nu
    =-\mathrm{d} t^2+\delta_{ij}\mathrm{d} x^i\mathrm{d} x^j, \qquad
    \bar{\phi}=qt,
    \end{equation}
as an exact solution if the function~$F_0(X)$ satisfies $F_0(-q^2)=F_{0X}(-q^2)=0$~\cite{Takahashi:2020hso}.
Assuming the existence of the Einstein-Hilbert Lagrangian [i.e., $F_2(X)\supset \MPl^2/2$] and that all the other terms in ${\cal L}_{\rm DHOST}$ are scaled by $\Lambda$, it is straightforward to study the dynamics of linear scalar perturbations about this background.
Due to the violation of the degeneracy, the dispersion relation acquires an $\omega^4$~term, which manifests the existence of the Ostrogradsky mode.
Expanding the two branches of solution for $\omega^2$ into a series with respect to $k/\Lambda$, we have~\cite{Motohashi:2019ymr,DeFelice:2022xvq}
\begin{align}
    \frac{\omega_1^2}{\Lambda^2}\simeq \tilde{s}_2\frac{\Lambda^2}{\MPl^2}\frac{k^2}{\Lambda^2}+\alpha\tilde{s}_4\frac{k^4}{\Lambda^4}+{\cal O}\left(\frac{k^6}{\Lambda^6}\right), \qquad
    \frac{\omega_2^2}{\Lambda^2}\simeq \frac{p}{\alpha}+{\cal O}\left(\frac{k^2}{\Lambda^2}\right), \label{disp_scordatura}
\end{align}
where the coefficients~$\tilde{s}_2$, $\tilde{s}_4$, and $p$ are dimensionless constants of order unity and we displayed the dependence on $\alpha$ when necessary.
The second branch diverges in the limit of $\alpha \to 0$, which is a typical behavior of the Ostrogradsky mode.
Since this mode does not satisfy the condition~$\omega_{2}^{2} \ll \Lambda^{2}$, it is beyond the regime of validity of the EFT.
On the other hand, the first branch is the physical mode.
Although the $k^2$~term is suppressed by $\Lambda^2/\MPl^2$, the ($k$-dependent) sound speed can be of order unity all the way up to the cutoff scale~$\Lambda$ thanks to the $k^4$~term introduced by detuning the degeneracy conditions.
Of course, in order to have a real sound speed, we need to require the positivity of the $k^4$~term, i.e., $\alpha\tilde{s}_4>0$.
Note also that, as it should be, the dispersion relation for the physical mode reduces to the form of \eqref{disp_EFT} in the limit of $\Lambda/\MPl\to 0$.
These facts show that the scordatura effect works successfully for the model~\eqref{scordatura}.

In general, the scordatura effect is omnipresent in nondegenerate higher-order scalar-tensor theories provided that the deviation from general relativity is controlled by a single mass scale which is well below $\MPl$.\footnote{Recently, an EFT of black hole perturbations with a time-dependent scalar field was constructed, which applies to both stealth and non-stealth solutions~\cite{Mukohyama:2022enj}.
When the EFT is applied to a stealth solution, it naturally accommodates the scordatura mechanism.}
Interestingly, the authors of \cite{DeFelice:2022xvq} showed that the scordatura effect is built-in in so-called U-DHOST theories, in which the degeneracy conditions~\eqref{eq:degeneracycondition} are partially broken in such a way that higher-derivative terms are degenerate {\it only} in the unitary gauge.
In this case, the Ostrogradsky mode [i.e., the one corresponding to the second branch in \eqref{disp_scordatura}] is intrinsically absent, but instead there is a nondynamical mode living on a spacelike hypersurface (i.e., the ``shadowy'' mode) which requires a careful treatment~\cite{DeFelice:2018ewo,DeFelice:2021hps}.

\section{Algorithm for characteristic analysis}
\label{app:algorithmcurves}

We review an algorithm for the characteristic analysis introduced in \cite{Motloch:2016msa} and apply it to our problem as a demonstration.
For the monopole perturbation about the stealth Schwarzschild solutions, as we saw in Sec.~\ref{ssec:monopole_master}, one can derive the quadratic Lagrangian in terms of a single master variable, from which the characteristic curves as well as the sound speed can be read off.
As it should be, by applying the algorithm, one just reproduces the results obtained directly from the quadratic Lagrangian.
Nevertheless, we present the algorithm for future reference because it would be useful in a more general setup (e.g., scordatura DHOST theories) where master variables are not always available.
In such a case, the algorithm serves as a powerful tool since it does not require the knowledge of master variables.

The algorithm is based on the ``Kronecker canonical form" (KCF) of a matrix pencil. 
Suppose we have a Lagrangian defined in a two-dimensional spacetime, where the time and spatial coordinates are $\tau$ and $\rho$, respectively.
We derive the equations of motion from the Lagrangian and rewrite them into the first-order form:
\begin{align}
    \mathbb{A} \dot{\bf{u}} + \mathbb{B} \bf{u}^{\prime} + \mathbb{C} \bf{u} = 0,
    \label{eq:1stderequ}
\end{align}
where $\mathbb{A}$, $\mathbb{B}$, $\mathbb{C}$ are $M \times N$ coefficient matrices and $\bf{u}$ is an $N$-dimensional vector. 
Here, a dot and a prime denote the derivative with respect to $\tau$ and $\rho$, respectively.
Next, we construct a matrix pencil $\mathbb{A}+\lambda \mathbb{B}$, where $\lambda$ is a label to keep track the spatial derivative. 
We perform the following transformations: \begin{align}
    \bf{v} = \mathbb{Q}^{-1} \bf{u}, \qquad
    \tilde {\mathbb{A}} = \mathbb{P A Q}, \qquad
    \tilde {\mathbb{B}} = \mathbb{P B Q}, \qquad
    \tilde {\mathbb{C}} = \mathbb{PCQ} + \mathbb{PA \dot{Q}} + \mathbb{PB Q^{\prime}},
\end{align}
with $\mathbb{P}$ and $\mathbb{Q}$ being invertible matrices having the size of $M\times M$ and $N\times N$, respectively.
These transformations amount to a change of variables and a linear combination of the equations of motion, and hence Eq.~\eqref{eq:1stderequ} can be rewritten as
\begin{align}
    \tilde{\mathbb{A}} \dot{\bf{v}} + \tilde{\mathbb{B}} \bf{v}^{\prime} + \tilde{\mathbb{C}} \bf{v} = 0.
\end{align}
For appropriate choices of $\mathbb{P}$ and $\mathbb{Q}$, the matrix pencil~$\tilde{\mathbb{A}}+\lambda \tilde{\mathbb{B}}$ can be brought into the KCF:
\begin{align}
    \tilde{\mathbb{A}}+\lambda \tilde{\mathbb{B}} 
    = 
    \mathrm{diag}(\mathbb{L}_{m_{1}},\cdots,\mathbb{L}_{m_{u}},\mathbb{L}^{P}_{n_{1}},\cdots,\mathbb{L}^{P}_{n_{o}},\mathbb{R}_{k_{1}}(\kappa_{1}),\cdots,\mathbb{R}_{k_{r}}(\kappa_{r})).
\end{align}
Here, $\mathbb{L}_{m}$ is an $m \times (m+1)$ bidiagonal matrix pencil, $\mathbb{L}^{P}_{n}$ is its pertransposition, 
and $\mathbb{R}_{k}(\kappa)$ is a $k \times k$ regular matrix pencil, defined by
\begin{equation}
\begin{split}
    &\mathbb{L}_{m} =
    \begin{pmatrix}
      0 & 1      &        & \\
        & \ddots & \ddots & \\
        &        &   0    & 1
    \end{pmatrix}
    + \lambda
    \begin{pmatrix}
      1 & 0      &         & \\
        & \ddots & \ddots  & \\
        &        &   1     & 0
    \end{pmatrix},
    \qquad
    \mathbb{L}^{P}_{m} =
    \begin{pmatrix}
      1 &        &   \\
      0 & \ddots &   \\
        & \ddots & 1 \\
        &        & 0
    \end{pmatrix}
    + \lambda
    \begin{pmatrix}
      0 &        &   \\
      1 & \ddots &   \\
        & \ddots & 0 \\
        &        & 1
    \end{pmatrix},
    \\
    &\mathbb{R}_{k}(\kappa) =
    \begin{cases}
      \lambda \mathbb{E}_{k}
      -
      \begin{pmatrix}
        \kappa &        &        & \\  
        1      & \kappa &        & \\
               & \ddots & \ddots & \\
               &        &   1    & \kappa  
      \end{pmatrix} 
      \,\,\mathrm{for}\,\,\, |\kappa| < \infty,
      \\
      \begin{pmatrix}
        \kappa &        &        & \\  
        1      & \kappa &        & \\
               & \ddots & \ddots & \\
               &        &   1    & \kappa  
      \end{pmatrix} 
      - \mathbb{E}_{k}
      \,\,\mathrm{for}\,\,\, |\kappa| = \infty,
    \end{cases}
\end{split}
\end{equation}
respectively, where $\mathbb{E}_{k}$ is a $k \times k$ identity matrix. 

The characteristic curves of the fields are determined by the regular blocks in the KCF of the matrix pencil. 
For the fields associated with a regular block~$\mathbb{R}_{k}(\kappa)$ with $\kappa$ being real, the slope of the characteristic curves is defined by $\kappa$~\cite{Motloch:2016msa}, i.e.,
\begin{align}
    \frac{\mathrm{d}\tau}{\mathrm{d}\rho} = -\kappa. 
    \label{eq:propagatedirection}
\end{align}
Therefore, the fields propagate along this direction.

\subsection{Example: massless monopole test scalar field in the Schwarzschild spacetime}

As an example, let us consider the following Lagrangian describing a massless monopole scalar field $\Phi$ in the Schwarzschild spacetime:
\begin{align}
    {\cal L} = \frac{1}{2} (\partial_{\mu} \Phi)^{2}, \qquad
    \Phi = \Phi(\tau, \rho),
\end{align}
where $\tau$ and $\rho$ are the Lema\^itre coordinates.
The equation of motion of the scalar field is given by
\begin{align}
    \ddot{\Phi} - \frac{1}{1-A(r)} \Phi^{\prime \prime} - \bra{ \frac{1}{1-A(r)} }^{\prime} \Phi^{\prime}  = 0.
\end{align}
In order to reduce the equation of motion into the first-order form, 
we introduce two auxiliary fields
\begin{align}
    \Phi_{\tau} = \dot \Phi, \qquad
    \Phi_{\rho} = \Phi^{\prime}. 
\end{align}
Note that these auxiliary fields satisfy
$\dot \Phi_{\rho} = \Phi_{\tau}^{\prime}$.
Now, the matrix pencil is given by
\begin{align}
    \mathbb{A} + \lambda \mathbb{B} =
    \begin{pmatrix}
     1       && 0      && 0 \\
     \lambda && 0      && 0\\
     0       && 1      && - \frac{\lambda}{1-A} \\
     0       && \lambda && -1
    \end{pmatrix}.
\end{align}
Then, the KCF of the matrix pencil is given by
\begin{align}
    \tilde{\mathbb{{A}}} + \lambda \tilde{\mathbb{B}}
    =
    \begin{pmatrix}
      1       && 0                    && 0 \\
     \lambda  && 0                    && 0\\
     0        && \lambda + \sqrt{1-A} && 0 \\
     0        && 0                    && \lambda - \sqrt{1-A}
    \end{pmatrix}
    = 
    \mathrm{diag}\bigl( \mathbb{L}_{1}, \mathbb{R}_{1}(\sqrt{1-A}), \mathbb{R}_{1}(-\sqrt{1-A}) \bigr). 
\end{align}

This example tells us the relation between the squared sound speed and the regular part of the KCF. 
When the regular part of a matrix pencil of the KCF is given by $\mathrm{diag}\bigl(\mathbb{R}_{1}(\kappa),\mathbb{R}_{1}(-\kappa) \bigr)$, 
we define the squared sound speed as~\cite{Takahashi:2021bml}
\begin{align}
    c_{s}^{2} = \frac{g_{\rho \rho}}{|g_{\tau \tau}|} \kappa^{-2}. 
\end{align}
For a massless monopole scalar field, the squared sound speed reads
\begin{align}
    c_{s}^{2} = \frac{g_{\rho \rho}}{|g_{\tau \tau}|} \frac{1}{1-A} = 1,
\end{align}
which corresponds to the (squared) speed of light. 
Therefore, a massless monopole scalar field propagates with the speed of light. 

From Eq.~\eqref{eq:propagatedirection}, we can read off the characteristic curves of massless monopole scalar fields as follows:  
\begin{align}
    \frac{\mathrm{d}\tau}{\mathrm{d}\rho} = \pm \sqrt{1-A}. 
\end{align}
Recasting this equation in terms of the usual Schwarzschild coordinates $\{t,r,\theta,\varphi\}$, we find
\begin{align}
    \frac{\mathrm{d}t}{\mathrm{d}r} = \pm \frac{1}{A}. 
\end{align}
This equation gives two curves in the $(t,r)$-plane, which lie in the null directions in the Schwarzschild spacetime. Therefore, a massless monopole scalar field propagates in the null directions.

\subsection{Monopole perturbations in DHOST theories}
\label{sssec:mopolemode}

Here, we apply the above algorithm to the case of the monopole perturbation in DHOST theories. 
Let us begin with the quadratic Lagrangian for the monopole perturbations [see Eq.~\eqref{eq:actionzeta}]
\begin{align}
    {\cal L}^{(2)}
  =
    \frac{d_{1}}{2} \dot \zeta^{2}
  - \frac{d_{2}}{2} \zeta ^{\prime 2}
  - \frac{d_{3}}{2} \zeta ^{2}.
\end{align}
From this Lagrangian, the equation of motion is given by
\begin{align}
  d_{1} \ddot{\zeta} - d_{2} \zeta^{\prime \prime} - d_{2}^{\prime} \zeta^{\prime} + d_{3}\zeta = 0.
\end{align}
To reduce the equation of motion into the first-order form, 
we introduce auxiliary fields as $\zeta_{\tau} = \dot \zeta$ and $\zeta_{\rho}=\zeta^{\prime}$. 
These auxiliary fields satisfy $\zeta_{\tau}^{\prime} = \dot{\zeta_{\rho}}$. 
Then, the KCF of the matrix pencil can be obtained as
\begin{align}
        \tilde{\mathbb{{A}}} + \lambda \tilde{\mathbb{B}}
    =
    \begin{pmatrix}
      1       && 0                    && 0 \\
     \lambda  && 0                    && 0\\
     0        && \lambda + \sqrt{\frac{d_{1}}{d_{2}}} && 0 \\
     0        && 0                    && \lambda - \sqrt{\frac{d_{1}}{d_{2}}}
    \end{pmatrix}
    = 
    \mathrm{diag}\bra{ \mathbb{L}_{1}, \mathbb{R}_{1}\bra{\sqrt{\frac{d_{1}}{d_{2}}}}, \mathbb{R}_{1}\bra{-\sqrt{\frac{d_{1}}{d_{2}}}} }. 
\end{align}
Therefore, the squared sound speed is given by
\begin{align}
    c_{s}^{2} = \frac{g_{\rho \rho}}{|g_{\tau \tau}|} \kappa^{-2} = (1-A) \frac{d_{2}}{d_{1}},
\end{align}
which coincides with Eq.~\eqref{eq:soundspeedDHOST}.

The characteristic curves are determined by the following equation:
\begin{align}
    \frac{\mathrm{d}t}{\mathrm{d}r}
    = - \frac{\sqrt{d_{1}} \pm \sqrt{d_{2}}(1-A)}{A\sqrt{1-A}(\sqrt{d_{1}} \pm \sqrt{d_{2}})},
    \label{eq:charactercurvesmono}
\end{align}
which can be integrated to yield
\begin{align}
    t 
    + 
    \frac{2}{3}\frac{r^{3/2}}{\sqrt{r_{\mathrm{s}}}}
    + 
    2 r_{\mathrm{g}} \sqrt{\frac{r}{r_{\mathrm{s}}}}
    +
    2 \sqrt{r_{\mathrm{s}}r}
    + 
    \frac{r_{\mathrm{g}}^{3/2}}{\sqrt{r_{\mathrm{s}}}} \ln \norm{ \frac{\sqrt{r} - \sqrt{r_{\mathrm{g}}}}{\sqrt{r} + \sqrt{r_{\mathrm{g}}}} }
    + 
    r_{\mathrm{s}} \ln 
    \left| \frac{\sqrt{r} - \sqrt{r_{\mathrm{s}}}}{\sqrt{r} + \sqrt{r_{\mathrm{s}}}} \right|
    = \mathrm{const},
    \\
    t 
    + 
    \frac{2}{3}\frac{r^{3/2}}{\sqrt{r_{\mathrm{s}}}}
    -
    2 r_{\mathrm{g}}\sqrt{\frac{r}{r_{\mathrm{s}}}}
    + 
    2 \sqrt{r_{\mathrm{s}} r}
    + \frac{2 r_{\mathrm{g}}^{3/2}}{\sqrt{r_{\mathrm{s}}}} \tan^{-1} \bra{ \sqrt{\frac{r}{r_{\mathrm{g}}}} }
    + 
    r_{\mathrm{s}} \ln 
    \left| \frac{\sqrt{r} - \sqrt{r_{\mathrm{s}}}}{\sqrt{r} + \sqrt{r_{\mathrm{s}}}} \right|
    = \mathrm{const},
\end{align}
with $r_{\mathrm{g}} \equiv \sqrt{D_{2}/D_{1}}$. 
These expressions are nothing but those of $\tilde{v}$ and $\tilde{u}$, respectively [see Eqs.~\eqref{eq:directionnomopolev} and~\eqref{eq:directionnomopoleu}]. 
Note that the characteristic curves of the monopole perturbations in the DHOST theory do not coincide with those of null rays in the Schwarzschild spacetime.

\section{Choice of initial surface of two-dimensional wave equation}
\label{app:choiceinitialsurface}

The equation of motion of the black hole perturbation can be often written in the form of a two-dimensional wave equation: 
\begin{align}
    \frac{\partial^{2} Q}{\partial x_{2}^{2}} - \frac{\partial^{2} Q}{\partial x_{1}^{2}} - {\cal V}_{\mathrm{eff}} (x_{2}) Q = 0,
    \label{eq:mastereq1}
\end{align}
where $x_{1}$ and $x_{2}$ are spacetime coordinates, $Q$ is a master variable, and ${\cal V}_{\mathrm{eff}}$ is an effective potential, which is a function of $x_{2}$. 
When we try to determine the evolution of the perturbation as the Cauchy problem, we should specify the initial surface. 
Naively, there are two physically different ways to solve the two-dimensional wave equation~\eqref{eq:mastereq1}
by imposing initial conditions on either $x_{1}={\rm const.}$~surface or $x_{2}={\rm const.}$~surface. 
Depending on the physical situation, we should carefully choose an appropriate initial surface.
\begin{figure}[t]
\includegraphics[width=100mm]{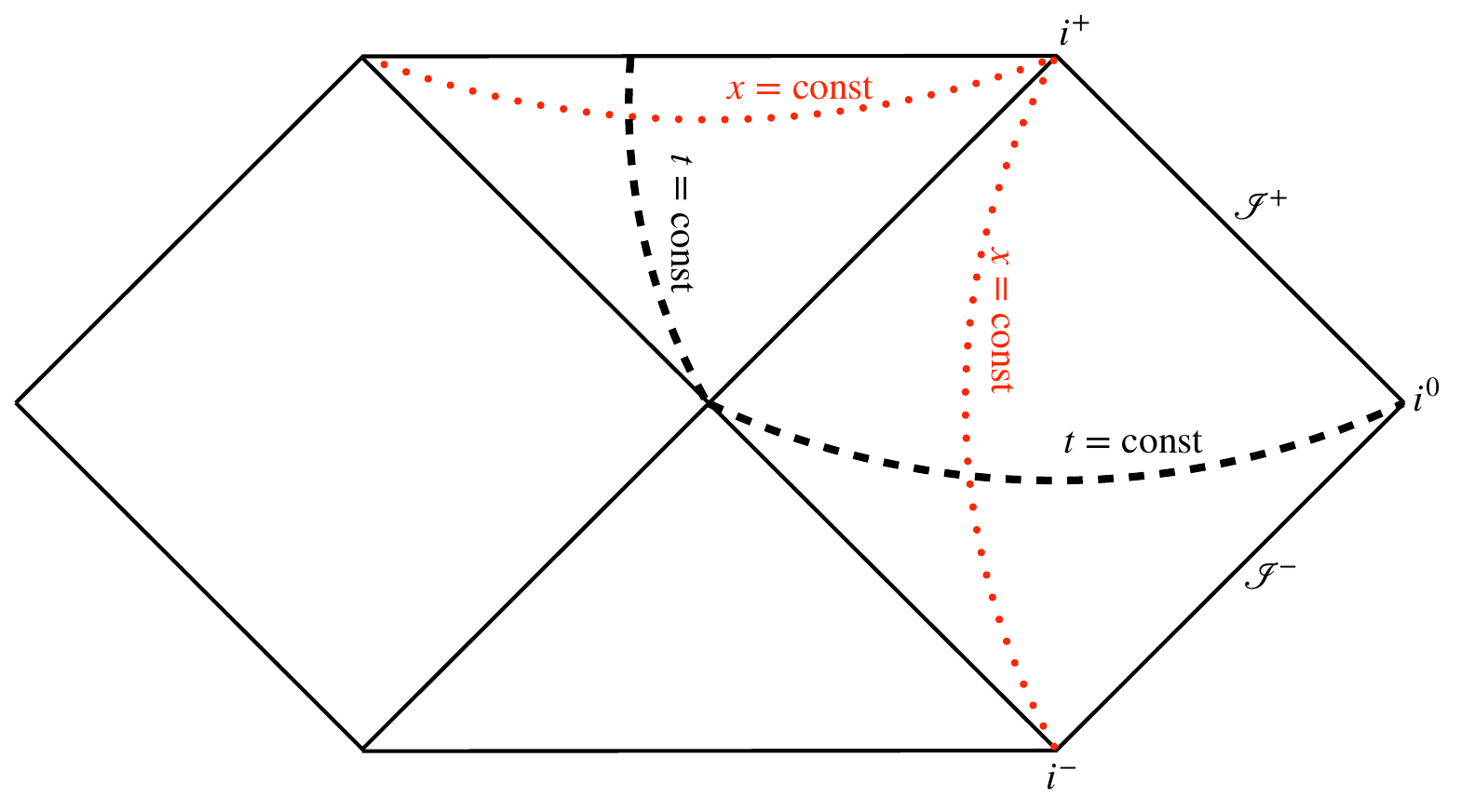}
\caption{Black dashed curves are $t = \mathrm{const.}$~surfaces and red dotted curves are $x=\mathrm{const.}$~surfaces. }
\label{fig:ambiguity}
\end{figure}

As an example, we consider the massless test scalar field in the Schwarzschild spacetime. 
After decomposing the scalar field with the spherical harmonics~$Y_{\ell m}(\theta,\phi)$, the equation of motion for the expansion coefficients~$u_{\ell m}$ can be expressed in the form of Eq.~\eqref{eq:mastereq1} as follows: 
\begin{align}
    \frac{\partial^{2} u_{\ell m}}{\partial x^{2}}-\frac{\partial^{2} u_{\ell m}}{\partial t^{2}} - {\cal V_{\ell}}(x) u_{\ell m} = 0,
    \label{eq:masslesseom}
\end{align}
where ${\cal V_{\ell}}$ is the effective potential for the massless scalar field, $t$ is the Killing time, and $x$ is the standard tortoise coordinate defined by $x \equiv r + r_{\mathrm{s}}\ln |r/r_{\mathrm{s}}-1|$. 
Comparing Eqs.~\eqref{eq:mastereq1} and~\eqref{eq:masslesseom}, 
we can read that $x_{1}$ and $x_{2}$ correspond to $t$ and $x$, respectively. 
Figure~\ref{fig:ambiguity} shows both $t=\mathrm{const}.$~surface (black  dashed curve) and $x=\mathrm{const.}$~surface (red dotted curve) in the Penrose diagram. 
When we determine the evolution of the scalar field in the exterior of the event horizon, the physically appropriate initial surface is a $t=\mathrm{const}.$~surface. 
However, if we determine the evolution of the scalar field in the interior of the event horizon, we should choose an $x=\mathrm{const}.$~surface as the initial surface because it is a spacelike hypersurface inside the event horizon.

In the DHOST theory we considered in the main text, the equation of motion for the monopole perturbations about stealth Schwarzschild solutions is expressed as Eq.~\eqref{eq:Shcroeq}. 
Comparing Eqs.~\eqref{eq:mastereq1} and~\eqref{eq:Shcroeq}, $x_{1}$ and $x_{2}$ correspond to $\tilde{t}$ and $r_{*}$, respectively. 
Looking at the form of the equation, one may naively want to choose a $\tilde{t}=\mathrm{const.}$~surface as the initial surface.
However, a $\tilde{t}=\mathrm{const.}$~surface is not a spacelike hypersurface.
Therefore, in the main text we did not impose the initial conditions on a $\tilde{t}=\mathrm{const.}$~surface. 
Instead, in Sec.~\ref{sec:timeevolution}, we chose the spacelike hypersurface~$\Sigma$ as the initial surface and formulated a physically sensible initial value problem.

\section{Numerical method and convergence test}
\label{app:method}

Let us explain how we solve Eq.~\eqref{eq:doublenullequation} numerically. 
We note that, in this Appendix, we set $r_{\mathrm{s}}=1$ for simplicity. 
We employ the numerical method introduced in~\cite{Gundlach:1993tp,Lucietti:2012xr}. 
In this method, we divide the numerical domain by characteristic curves of the fields. 
In the DHOST theory, since the monopole perturbations propagate along 
the directions of constant $\tilde{U}$ and constant $\tilde{W}$, we divide the numerical domain with $\tilde{U}$ and $\tilde{W}$. 
We discretize the coordinates~$\tilde{U}$ and $\tilde{W}$ as $(\tilde{U}_{i},\tilde{W}_{j})$ with $i=0,1,2,\cdots,N$ and $j=0,1,2,\cdots,M$. Denoting $\Psi_{i,j}=\Psi(\tilde{U}_{i},\tilde{W}_{j})$, $\tilde{V}_{i,j}=\tilde{V}(\tilde{U}_{i},\tilde{W}_{j})$, and $h=\tilde{U}_{i+1}-\tilde{U}_{i}=\tilde{W}_{i+1}-\tilde{W}_{i}$, we can discretize Eq.~\eqref{eq:doublenullequation} as 
\begin{align}
    \Psi_{i+1,j+1} = 
      \Psi_{i,j+1}
     +\Psi_{i+1,j}
     -\Psi_{i,j}
     -\frac{h^{2}}{4}(\tilde{V}\Psi)_{i+1/2,j+1/2}
     +{\cal O}(h^{4}),
     \label{eq:descreteequation}
\end{align}
where
\begin{align}
    (\tilde{V}\Psi)_{i+1/2,j+1/2}
    = 
    \frac{1}{4}
    \bra{
       \tilde{V}_{i+1,j+1} \Psi_{i+1,j+1}
      +\tilde{V}_{i+1,j}\Psi_{i+1,j}
      +\tilde{V}_{i,j+1}\Psi_{i,j+1}
      +\tilde{V}_{i,j}\Psi_{i,j}
    }.
\end{align}
Solving Eq.~\eqref{eq:descreteequation} for $\Psi_{i+1,j+1}$, we have
\begin{align}
    \Psi_{i+1,j+1} = 
    \bra{
      1+\frac{h^{2}}{16} \tilde{V}_{i+1,j+1}
    }^{-1}
    \bra{
       \Psi_{i,j+1}
      +\Psi_{i+1,j}
      -\Psi_{i,j}
      -\frac{h^{2}}{4}
      (\tilde{V}\Psi)^{\prime}_{i+1/2,j+1/2}
    }
    +{\cal O}(h^{4}),
    \label{eq:discretizedfourpoint}
\end{align}
where 
\begin{align}
    (\tilde{V}\Psi)^{\prime}_{i+1/2,j+1/2} 
    = 
    \frac{1}{4}
    \bra{
      \tilde{V}_{i+1,j}\Psi_{i+1,j}
     +\tilde{V}_{i,j+1}\Psi_{i,j+1}
     +\tilde{V}_{i,j}\Psi_{i,j}
    }.
\end{align} 
Through Eq.~\eqref{eq:discretizedfourpoint}, $\Psi_{i+1,j+1}$ is calculated from $\Psi$ and $\tilde{V}$ at $(\tilde{U}_{i},\tilde{W}_{j})$, $(\tilde{U}_{i+1},\tilde{W}_{j})$, and $(\tilde{U}_{i},\tilde{W}_{j+1})$.

In order to obtain the solution in the numerical domain, we need to impose the initial conditions on the initial surface. We consider a Gaussian wave packet as the initial data: 
\begin{align}
    \Psi _{\rm ini} \equiv \Psi (\tilde{U},-\tilde{U}+\tilde{W}_{\mathrm{max}})
    = 
    \frac{1}{\sqrt{2\pi} \sigma} 
    e^{ -\frac{1}{2}\bra{
    {\frac{\tilde{U}-\tilde{U}_{0}}{ \sigma}}
    }^{2}},
    \label{eq:Gaussianinitialdata}
\end{align}
where $\tilde{W}_{\mathrm{max}}$ is the maximum value of the range for $\tilde{W}$, and $\sigma$ and $\tilde{U}_{0}$ define the typical width and central location of the initial Gaussian wave packet, respectively.
It should be noted that we truncated the Gaussian profile in a finite region in our actual computations.
We also note that we have determined the normalization factor so that $\int  \Psi_{\rm ini}\, \mathrm{d}\tilde{U} =1$. 
Thus, the peak value of the initial Gaussian wave packet is given by
\begin{align}
    \Psi_{\rm ini}^{\rm peak}\equiv \Psi_{\rm ini}|_{\tilde{U}=\tilde{U}_{0}}=
    \frac{1}{\sqrt{2\pi} \sigma}.
    \label{eq:psipeak}
\end{align}
Note that we set the value of $\tilde{U}_{0}$ so that the peak is located at the center of the region $S$. 
The normalization means that the amplitude of the longest-wavelength mode does not depend on the width of the initial Gaussian wave packet.
Indeed, the Fourier transformation of $\Psi_{\rm ini}$ is given by
\begin{align}
    \frac{1}{\sqrt{2\pi }}\int _{-\infty}^{\infty} \Psi_{\rm ini}\, e^{i \omega \tilde{U}} \mathrm{d} \tilde{U} =
    \frac{1}{\sqrt{2\pi}} e^{ i\omega \tilde{U}_{0} -  \omega^{2}\sigma^{2}/2}.
\end{align}
For $\omega=0$, the amplitude is always $1/\sqrt{2\pi}$ regardless of the value of $\sigma$.

In order to solve Eq.~\eqref{eq:doublenullequation}, 
we have to specify the values of not only $\Psi$ but also the derivative of $\Psi$ on the initial surface. 
We assume that the derivative in the direction perpendicular to $\Sigma$ vanishes on the initial surface~$\Sigma$. In the region~$S$, this condition reduces to $\partial_{\tilde{U}} \Psi + \partial_{\tilde{W}} \Psi|_{S} = 0$, which leads to
\begin{align}
    \Psi_{i,j} = \Psi_{i+1,j+1}
                 \brb{
                   1 + \frac{h^{3}}{24}
                     (\partial_{\tilde{U}} \Psi 
                   + \partial_{\tilde{W}} \Psi)
                 }
                 + {\cal O}(h^{4}),
\end{align}
on the initial surface~$\Sigma$. 
From this relation, we obtain the following equation:
\begin{align}
    \Psi_{i+1,j+1} = 
    \dfrac{\Psi_{i+1,j} + \Psi_{i,j+1}
     - h^{2}
     (  \tilde{V}_{i+1,j} \Psi_{i+1,j}
      + \tilde{V}_{i,j+1} \Psi_{i,j+1}
     )/16}{2 
     + h^{2}
     (
     \tilde{V}_{i,j}+\tilde{V}_{i+1,j+1}
     ) /16
     - h^{2}
     (
     \tilde{V}_{i+1,j}+\tilde{V}_{i,j+1}-2\tilde{V}_{i,j}
     )/24}
    + {\cal O} (h^{4}).
    \label{eq:derivativepsi}
\end{align}
Combining the discretized equation~\eqref{eq:discretizedfourpoint} and initial conditions~\eqref{eq:Gaussianinitialdata} and~\eqref{eq:derivativepsi}, we can obtain the solution~$\{ \Psi_{i,j} \}$ in the whole numerical domain.

Next, we clarify the convergence of the numerical method. 
To this end, we consider the absolute error and the relative error between the two solutions for $\Psi$ calculated with different resolutions as follows:
\begin{align}
    \Delta_{h_{2}-h_{1}} \equiv |\Psi_{h_2}-\Psi_{h_1}|,
    \,\,\,
    \delta_{h_{2}-h_{1}} \equiv 
    \left|
    \frac{\Psi_{h_2}-\Psi_{h_1}}{\Psi_{h_1}}\right|,
\end{align}
where $h_{1}$ and $h_{2}$ are two different grid sizes, which are ordered as $h_{1}<h_{2}$. 
\begin{figure}[t]
\includegraphics[width=150mm]{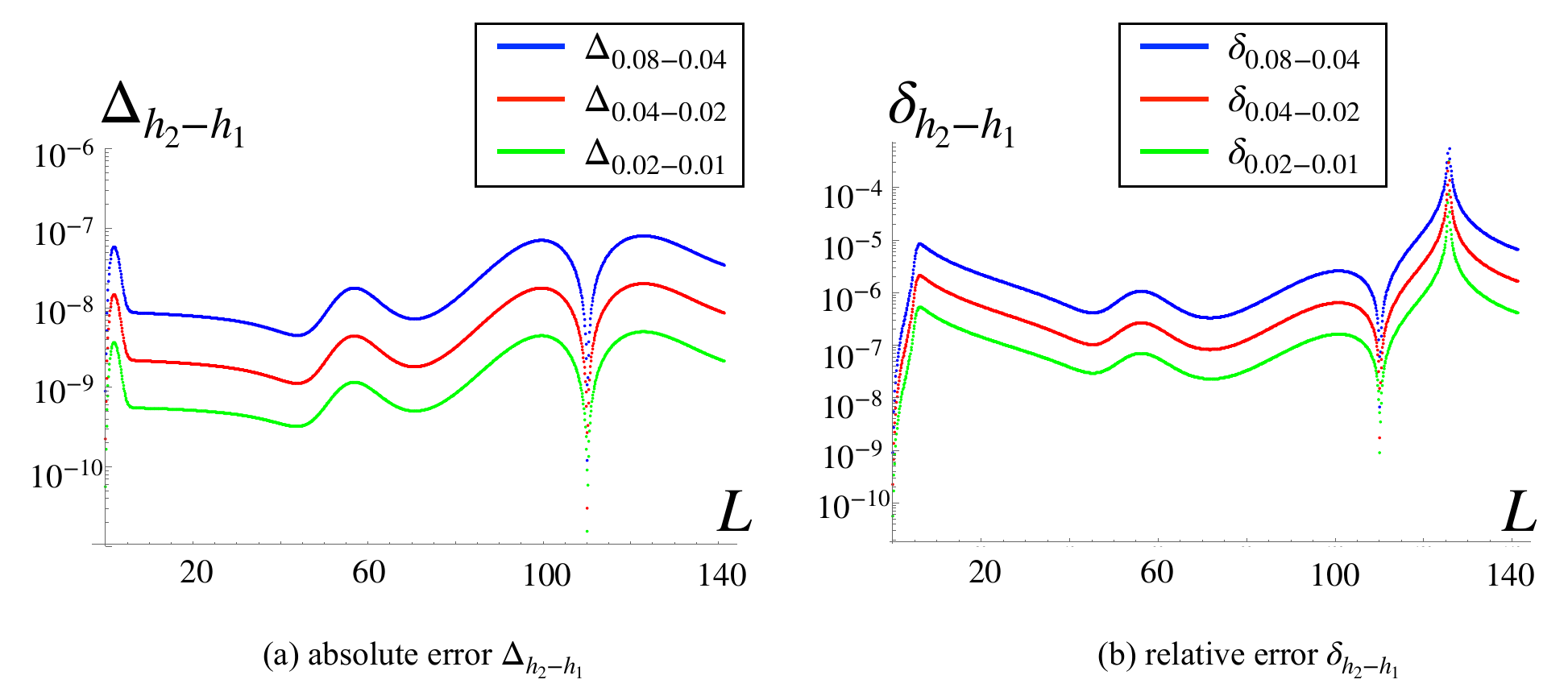}
\caption{The absolute error~$\Delta_{h_{2}-h_{1}}$ and relative error~$\delta_{h_{2}-h_{1}}$ along the line of $\tilde{U}=\tilde{W}$ calculated with different resolutions. 
The horizontal axis is the distance from the location of $\Psi_{\rm ini}^{\rm peak}$ in the $(\tilde{U},\tilde{W})$-space. Both the errors get quartered as the grid size is halved, showing the ${\cal O}(h^{2})$ convergence. 
}
\label{fig:error}
\end{figure}
Figure~\ref{fig:error} shows the absolute error and the relative error along the line of $\tilde{U}=\tilde{W}$.
Note that the horizontal axis in Fig.~\ref{fig:error} is the distance from the location of $\Psi_{\rm ini}^{\rm peak}$ in the $(\tilde{U},\tilde{W})$-space: $L \equiv \sqrt{2}(\tilde{U}-\tilde{U}_{0})$.
We also note that the relative error diverges at some $L \simeq 125$, but this does not mean the numerical calculation breaks down. 
The reason of the divergence is that the field value crosses zero at this point.
For the absolute error, we see that there is no divergence at $L \simeq 125$. 
According to Fig.~\ref{fig:error}, both the errors get quartered as the grid size is halved.
Thus, we confirm that our numerical method shows the ${\cal O}(h^{2})$ convergence. 
As we can see in the red dotted plot in Fig.~\ref{fig:error}(b), the maximum value of the relative error between $h=0.04$ and $h=0.02$ is about $0.1\%$. 
Since the numerical accuracy is sufficiently high for the grid size of $h=0.04$, we chose this value of $h$ in Sec.~\ref{sec:timeevolution}.

\section{\texorpdfstring{Numerical results for $r_{\mathrm{g}} \ge r_{\mathrm{s}}$}{Numerical results for rg is equal or greater than rs}}
\label{app:resultsother}

\begin{figure}[t]
\includegraphics[width=150mm]{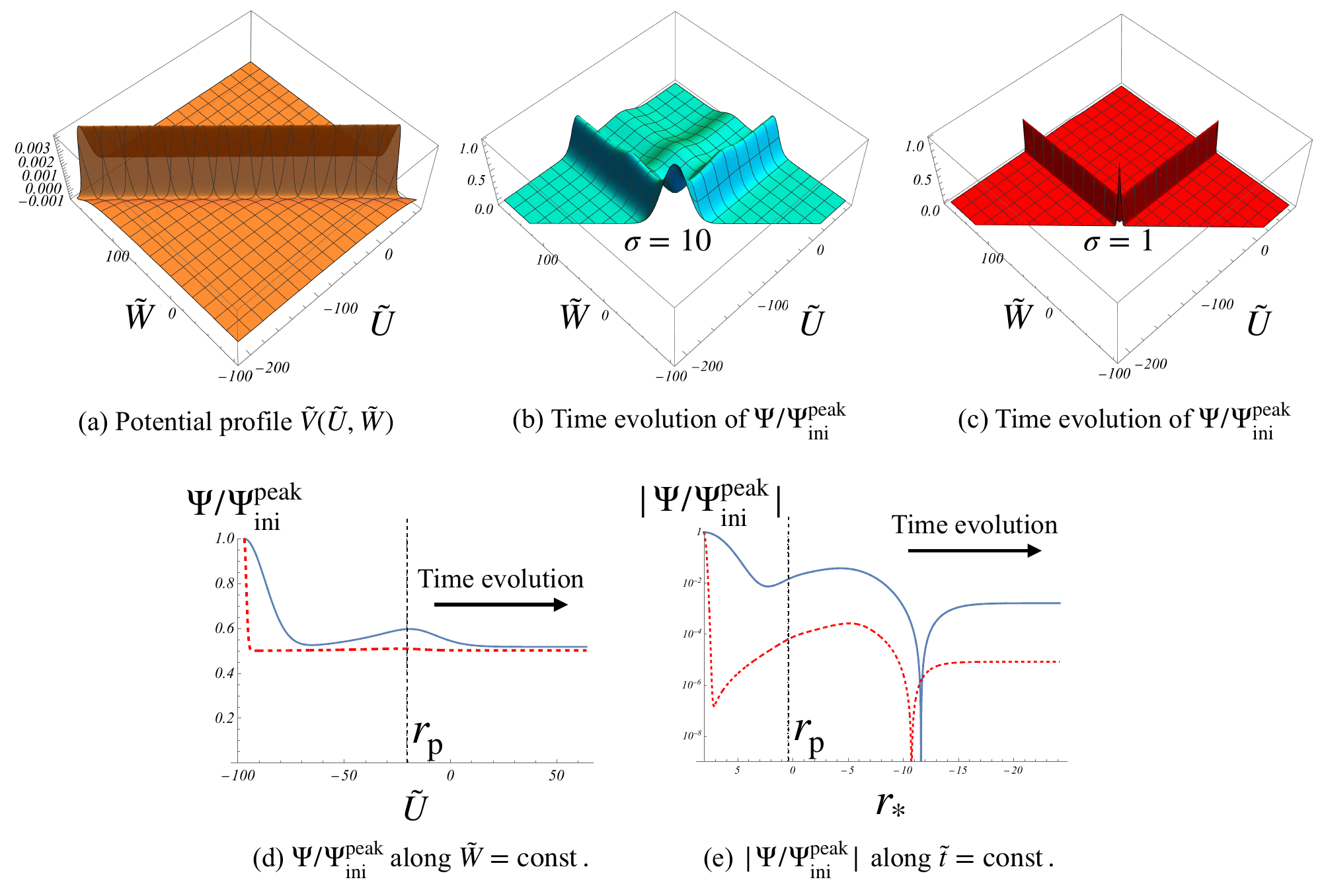}
\caption{Time evolution of the monopole perturbations for $r_{\mathrm{g}}=1$. 
In panels~(d) and (e), the blue solid curve and the red dashed curve are the time evolution of $\sigma=10$ and $\sigma=1$, respectively. The black dashed line corresponds to the radius of the photon sphere~$r_{\mathrm{p}}=1.5 r_{\mathrm{s}}$. }
\label{fig:timeevolution10}
\end{figure}

\begin{figure}[t]
\includegraphics[width=150mm]{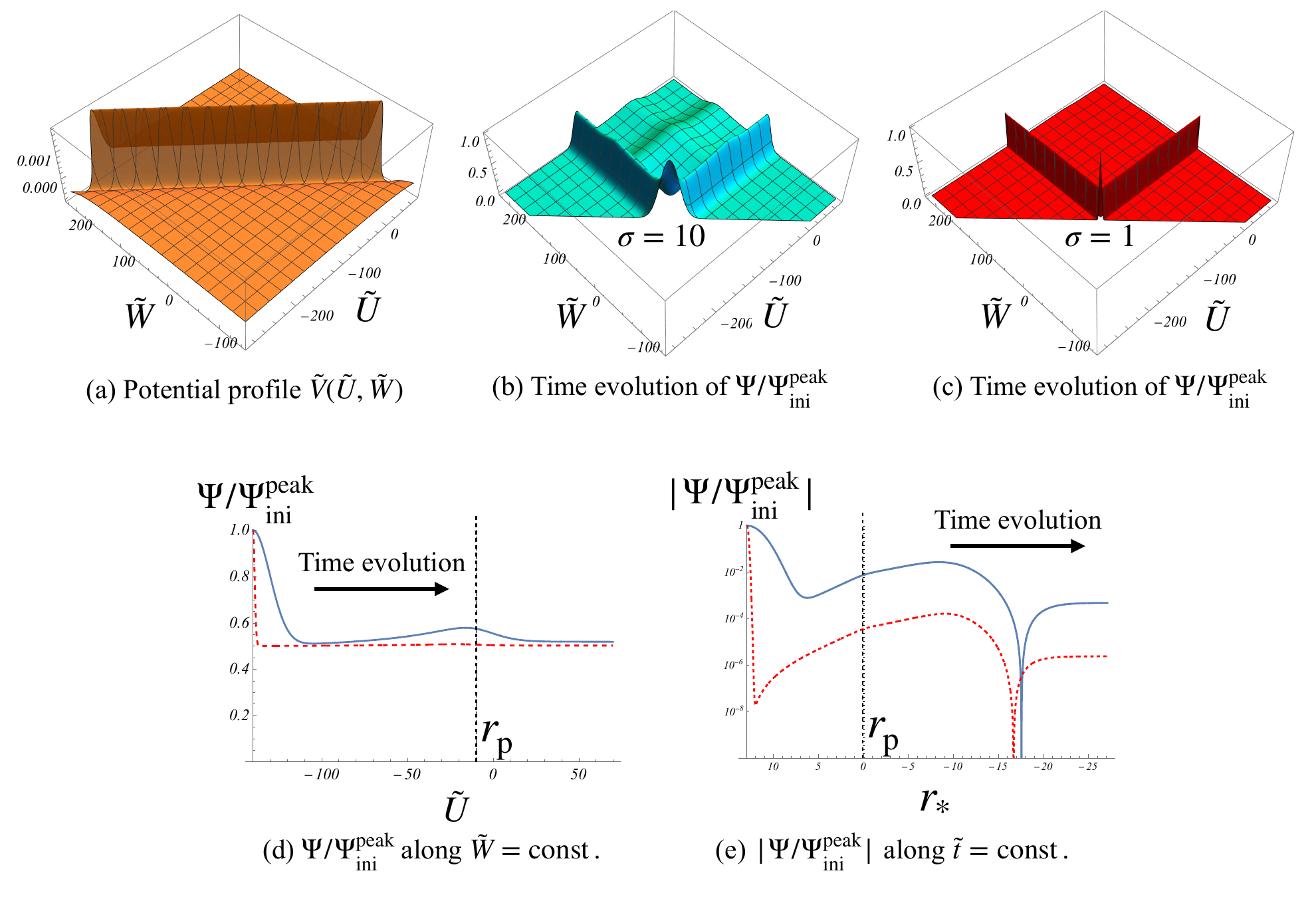}
\caption{Time evolution of the monopole perturbations for $r_{\mathrm{g}}=1.2$. 
In panels~(d) and (e), the blue solid curve and the red dashed curve are the time evolution of $\sigma=10$ and $\sigma=1$, respectively. The black dashed line corresponds to the radius of the photon sphere~$r_{\mathrm{p}}=1.5r_{\mathrm{s}}$. }
\label{fig:timeevolution12}
\end{figure}

As we mentioned in Sec.~\ref{subsec:tildetconstsurface}, depending on the choice of the theory parameters, the location of the monopole horizon~$r=r_{\mathrm{g}}$ changes, and there are three possibilities: $r_{\mathrm{g}}<r_{\mathrm{s}}$ ({\it case~I}\,), $r_{\mathrm{g}}=r_{\mathrm{s}}$ ({\it case~II}\,), and $r_{\mathrm{g}}>r_{\mathrm{s}}$ ({\it case~III}\,). 
In Sec.~\ref{subsec:scattering}, we focused on the {\it case~I} with $r_{\mathrm{g}} = 0.5 \,r_{\mathrm{s}}$.
As a complementary analysis, here we show the numerical results for the {\it case~II} with $r_{\mathrm{g}}=r_{\mathrm{s}}$ and the {\it case~III} with $r_{\mathrm{g}} = 1.2 r_{\mathrm{s}}$.
For simplicity, we set $r_{\mathrm{s}}=1$ for the following. Figures~\ref{fig:timeevolution10} and~\ref{fig:timeevolution12} show the time evolution of the monopole perturbations for $r_{\mathrm{g}}=1$ and $r_{\mathrm{g}}=1.2$, respectively. 
In these figures, the black dashed line depicts the position of the photon sphere~$r_{\mathrm{p}}=1.5r_{\mathrm{s}}$. 
For $r_{\mathrm{g}}=1$, the scaling constants~$a,b$ and the numerical domain are set to be $a=5$, $b=4.6$, $-100 \le \tilde{W} \le 176$, and $-212 \le \tilde{U} \le 64$, respectively. 
On the other hand, for $r_{\mathrm{g}}=1.2$, we set $a=5$, $b=4.7$, $-140 \le \tilde{W} \le 220$, and $-290 \le \tilde{U} \le 70$, respectively. For both cases, we choose the grid size as $h=0.04$.

In both cases, there is no qualitative difference from the $r_{\mathrm{g}}=0.5$~case studied in \ref{subsec:scattering}. 
The monopole perturbation grows when it propagates through the negative-potential region. 
When the monopole perturbation enters the positive-potential region, it begins to decay. 
However, since the potential decays exponentially, the monopole perturbation does not decay completely and one wave packet propagates towards the future timelike infinity~$i^{+}$ and the other towards the spacetime singularity. 
The damped oscillations at the late time do not show up.

As in the $r_{\mathrm{g}}=0.5$~case, there is still a possibility that the stealth Schwarzschild solutions in the DHOST theory are unstable for long-wavelength modes. 
If they are unstable for $r_{\mathrm{g}}\ge r_{\mathrm{s}}$, the instability would be serious, because the instability happens outside the Schwarzschild radius. 
On the other hand, if the instability shows up only for modes with sufficiently long wavelength, it would be harmless, similar to the Jeans instability.
Here, we conclude that, for all the three cases, the stealth Schwarzschild solutions in the DHOST theory are dynamically stable against the monopole perturbations with 
the wavelength comparable or shorter than the size of the black hole horizon.

\section{Growing phase in the negative-potential region}
\label{app:growingphase}

We analyze the growing phase appearing in the time evolution of the monopole perturbation in Sec.~\ref{subsec:scattering}.
The monopole perturbations obey the following equation:
    \begin{equation}
    \frac{\partial^2\Psi}{\partial \tilde{t}^2}-\frac{\partial^2\Psi}{\partial r_*^2}-ab\tilde{V}(r_*)=0,
    \end{equation}
where the functional form of the potential~$\tilde{V}$ is given in Eq.~\eqref{eq:Vtilde} (see also Fig.~\ref{fig:potentialrstar} for the potential profile).
In our set up, the monopole perturbation propagates from the negative-potential region to the positive-potential region. 
According to the numerical results, in the negative-potential region, which lies far outside the monopole horizon, the monopole perturbation grows.  
In order to estimate the growth rate, let us study the asymptotic behavior of the potential~$\tilde{V}$ as a function of the generalized tortoise coordinate~$r_*$.
For large $r$, Eq.~\eqref{eq:genetortoisecoordinate} can be expanded in the form, 
    \begin{equation}
    r_*=\frac{2r_{\rm g}^{3/2}}{r_{\rm s}^{1/2}}\bra{\frac{r}{r_{\rm g}}}^{1/2}
    +\rc+{\cal O}\bra{r^{-3/2}}, \label{eq:r*exp}
    \end{equation}
which can be solved for $r$ as
    \begin{equation}
    r=\frac{r_{\rm s}}{4r_{\rm g}^3}(r_*-\rc)^2+{\cal O}\bra{(r_*-\rc)^{-2}}.
    \end{equation}
The constant~$\rc$ is the integration constant originating from the integration in Eq.~\eqref{eq:genetortoisecoordinate_int}. 
Although we fixed $\rc$ to a particular finite value in Eq.~\eqref{eq:genetortoisecoordinate}, here we use $\rc$ to denote an arbitrary constant to keep track of the ambiguity of $r_*$.
Then, the potential~$\tilde{V}$ can be written as
    \begin{equation}
    ab\tilde{V}=-\frac{3}{4(r_*-\rc)^2}+{\cal O}\bra{(r_*-\rc)^{-6}}. \label{eq:Vtildeexp}
    \end{equation}
This would give a good approximation of $\tilde{V}$ for $r_*$ larger than $r_{*}^{\rm min}$, i.e., the location of the local minimum of the potential~$\tilde{V}$ (see Fig.~\ref{fig:potentialrstar}).
From Eq.~\eqref{eq:Vtilde}, in terms of the original radial coordinate~$r$, the local minimum of $\tilde{V}$ can be analytically obtained as
\begin{align}
    r(r_{*}^{\rm min}) = \sqrt{7+2\sqrt{\frac{23}{3}}} \,r_{\mathrm{g}} \sim 3.54\, r_{\mathrm{g}} ,
    \qquad
    ab\tilde{V}(r_{*}^{\rm min}) = 
    - \frac{9(19\sqrt{3}+7\sqrt{23})r_{\mathrm{s}}}{2(21+2\sqrt{69})^{5/2}r_{\mathrm{g}}^{3}} \sim - 0.034 \,\frac{r_{\mathrm{s}}}{r_{\mathrm{g}}^{3}},
\end{align}
from which we find
    \begin{equation}
    r_{*}^{\rm min}-\rc \sim 3.66\frac{r_{\rm g}^{3/2}}{r_{\rm s}^{1/2}}.
    \end{equation}
The above analysis shows that, it is not the coordinate value of $r_*$ itself but $r_*-\rc$ that has a physical meaning.
Therefore, without loss of generality, we fix $\rc=0$ in the following analysis.

Let us now consider a toy model
    \begin{equation}
    \frac{\partial^{2} \Psi}{\partial \tilde{t}^{2}} - \frac{\partial^{2} \Psi}{\partial r_{*}^{2}} - V_{\mathrm{toy}}(r_{*})\Psi = 0,
    \label{eq:waveeq_toy}
    \end{equation}
with the following toy potential, which would capture the large-$r_*$ behavior of $\tilde{V}$:
\begin{align}
    V_{\text{toy}}(r_{*})=
    \begin{cases}
     -\dfrac{m}{r_{*}^{2}} &(r_{*}^{\text{cut}}<r_{*}),\\
     0 &(r_{*} \le r_{*}^{\text{cut}}),
    \end{cases}
\end{align}
where $m$ is a positive constant parameter. 
Note that the asymptotic behavior of the potential~\eqref{eq:Vtildeexp} for the monopole perturbations amounts to $m=3/4$.
Here, we have introduced a cutoff~$r_{*}^{\text{cut}}\,(>0)$ to avoid the divergence of the field~$\Psi$ at $r_{*}=0$.
As mentioned above, the potential~$\tilde{V}$ would be well approximated by $\tilde{V}\propto -r_{*}^{-2}$ for $r_*>r_*^{\rm min}$, so we can choose $r_{*}^{\text{cut}}=r_*^{\rm min}$. 
Performing a Fourier transformation in $\tilde{t}$ 
such that $\Psi(\tilde{t},r_{*}) = \int \psi_\omega(r_{*}) e^{-i\omega \tilde{t}}\mathrm{d}\omega / \sqrt{2\pi}$, for $r>r_{*}^{\text{cut}}$, the two-dimensional wave equation~\eqref{eq:waveeq_toy} becomes 
\begin{align}
    \frac{\mathrm{d}^{2}\psi_\omega}{\mathrm{d}r_{*}^{2}} 
    + 
    \bra{
    \omega^{2} 
    - \frac{m}{r^{2}_{*}}
    } \psi_\omega = 0.
\end{align}
The general solution to this equation is given by
\begin{align}
    \psi_\omega(r_{*}) 
    = 
    {\cal A}\sqrt{r_{*}} N_{\sqrt{1+4m}/2} (|\omega| r_{*})
    +
    {\cal B}\sqrt{r_{*}} J_{\sqrt{1+4m}/2}(|\omega| r_{*}),
    \label{eq:psiomega}
\end{align}
where ${\cal A}$ and ${\cal B}$ are integration constants. 
Here, $N_{\alpha}(x)$ and $J_{\alpha}(x)$ are the Neumann function and the Bessel function of the first kind, respectively. 
The Neumann function~$N_{\alpha}(|\omega| r_{*})$ diverges at $r_{*}=0$, while the Bessel function~$J_{\alpha}(|\omega| r_{*})$ is finite for $r_{*} \ge 0$. 
For $m>0$, the general solution $\psi_\omega(r_*)$ always diverges at $r_{*}=0$ unless ${\cal B}=0$. 
In fact, $\psi_\omega$ can be expanded around $r_{*}=0$ as follows:
\begin{align}
    \psi_\omega \simeq
    \tilde{\cal A}
    \sqrt{r_{*}} \bra{|\omega| r_{*}}^{-\frac{\sqrt{1+4m}}{2}} 
    + 
    \tilde{\cal B}
    \sqrt{r_{*}}
    (|\omega|r_{*})^{\frac{\sqrt{1+4m}}{2}},
    \label{eq:seriesRstar0}
\end{align}
where $\tilde{\cal A}$ and $\tilde{\cal B}$ are given by 
\begin{align}
    \tilde{\cal A} = 
    - {\cal A}\, \frac{2^{\frac{\sqrt{1+4m}}{2}}}{\pi} \Gamma \bra{ \frac{\sqrt{1+4m}}{2}},
    \quad
    \tilde{\cal B} =
     {\cal B}\brb{2^{\frac{\sqrt{1+4m}}{2}}\,\Gamma \bra{ 1 + \frac{\sqrt{1+4m}}{2}}}^{-1},
\end{align}
with $\Gamma(\alpha)$ being the Gamma function. 
The first term in the right-hand side of Eq.~\eqref{eq:seriesRstar0} diverges in the limit~$r_{*}\to 0$ as long as $m>0$. 
Therefore, for a smaller value of $r_{*}^{\mathrm{cut}}$, the field~$\psi_\omega$ can grow larger.

We also discuss the $\omega$-dependence of the growth rate of the field from $r_{*}=r_{*}^{\rm ini}$ to $r_{*}=r_{*}^{\rm fin}$, with $0<r_{*}^{\rm fin}\ll r_{*}^{\rm ini}$. 
Let us focus on modes with $r_{*}^{\rm fin}\ll |\omega|^{-1}\ll r_{*}^{\rm ini}$. 
For simplicity, we keep only the growing mode, i.e., 
\begin{align} \label{eq:grow}
\psi_\omega^{\rm grow} = {\cal A}\sqrt{r_{*}} N_{\sqrt{1+4m}/2} (\omega r_{*}). 
\end{align}
Since the growing mode~$\psi_{\omega}^{\rm grow}$ oscillates in the asymptotic region, it is convenient to consider the envelope of $\psi_{\omega}^{\rm grow}$ to analyze the $\omega$-dependence of the growth rate. 
As is well-known, the envelope function of the Neumann function~$N_\alpha(x)$ or the Bessel function~$J_\alpha(x)$ is given by $[N^2_\alpha(x) + J^2_\alpha(x)]^{1/2}$~\cite[\S 10.18]{NIST:DLMF}.
Therefore, the envelope function of $\psi_{\omega}^{\rm grow}$ is given by 
\begin{align}
    {\cal E}^{\rm grow}(r_*)= 
    {\cal A} \sqrt{r_{*} \brb{N_{\sqrt{1+4m}/2}^{2}(\omega r_{*}) + J_{\sqrt{1+4m}/2}^{2}(\omega r_{*})}}. 
\end{align}
\begin{figure}[t]
\includegraphics[width=\textwidth]{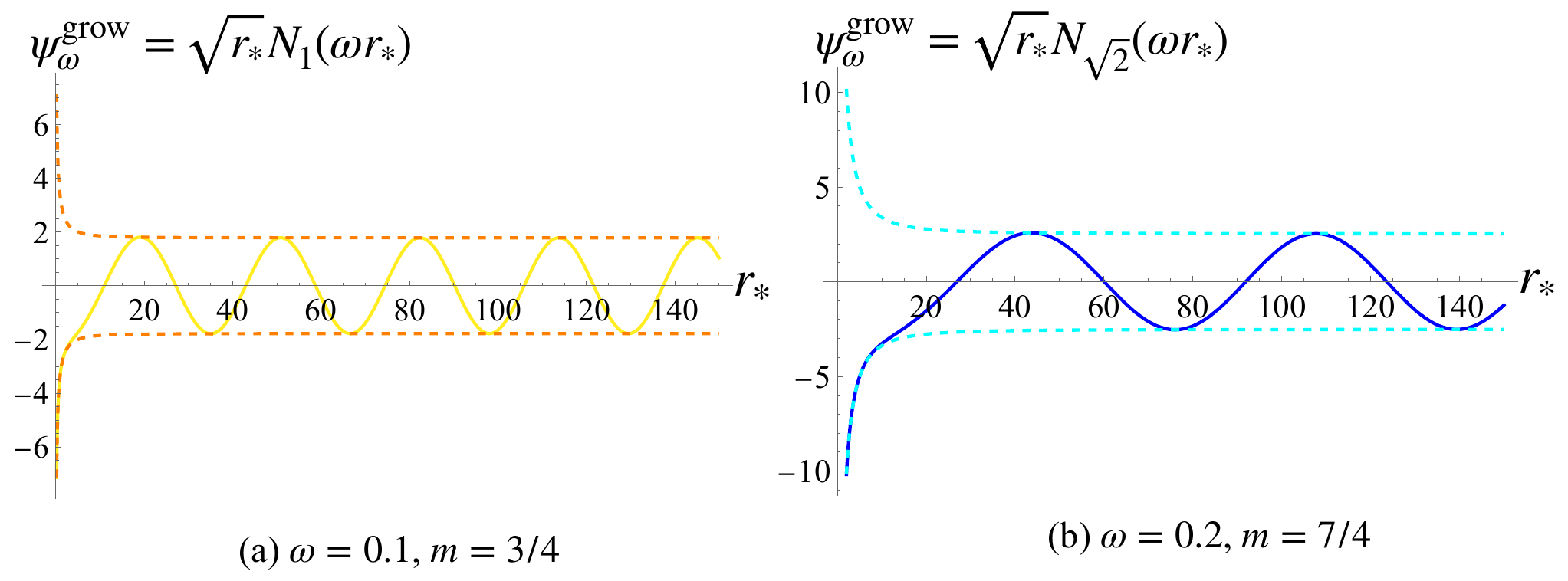}
\caption{The growing mode~$\psi_{\omega}^{\rm grow}$ (yellow/blue solid curves) and the envelope function~${\cal E}^{\rm grow}(r_{*})$ (orange/cyan dashed curves) for some specific values of $\omega$ and $m$: (a) $\omega=0.1$, $m=3/4$ and (b) $\omega=0.2$, $m=7/4$. Here, we set ${\cal A}=1$. }
\label{fig:envelope}
\end{figure}
Figure~\ref{fig:envelope} shows the growing mode $\psi_{\omega}^{\rm grow}$ (yellow/blue solid curves) and the envelope function (orange/cyan dashed curves) for some specific values of $\omega$ and $m$. 
At the initial value~$r_{*}=r_{*}^{\rm ini}$, 
the envelope function behaves as 
\begin{align}
  {\cal E}^{\rm grow}(r_*^{\rm ini}) \simeq
  {\cal A}\sqrt{\frac{2}{\pi|\omega|}}.
  \label{eq:envelopelargeeq}
\end{align}
This is consistent with the asymptotic form of the growing mode~\eqref{eq:grow} for $\omega r_{*} \gg 1$, which is given by
\begin{equation}
    \psi_{\omega}^{\rm grow}(r_{*})\simeq 
{\cal E}^{\rm grow}(r_*^{\rm ini})    
    \sin\bra{|\omega| r_{*}-\frac{1+\sqrt{1+4m}}{4}\pi}.
\end{equation}
At the final value~$r_* = r_{*}^{\rm fin}$, 
the growing mode $\psi_\omega^{\rm grow}$ is well approximated by 
the first term in Eq.~\eqref{eq:seriesRstar0}. 
Hence, the envelope function behaves as 
\begin{align}
    {\cal E}^{\rm grow}(r_*^{\rm fin}) \simeq 
    \frac{{\cal A}}{\pi} \Gamma\bra{\frac{\sqrt{1+4m}}{2}}
    \sqrt{\frac{2}{|\omega|}}\bra{\frac{|\omega| r_{*}^{\rm fin}}{2}}^{-\frac{1}{2}(\sqrt{1+4m}-1)}
    \simeq |\psi_\omega^{\rm grow}(r_*^{\rm fin})|.
\end{align}
Then, the growth rate can be estimated by
the ratio of ${\cal E}^{\rm grow}(r_*^{\rm fin})$ to ${\cal E}^{\rm grow}(r_*^{\rm ini})$, which is given by 
\begin{align}
    \frac{ {\cal E}^{\rm grow}(r_*^{\rm fin}) }{ {\cal E}^{\rm grow}(r_*^{\rm ini}) }
    \simeq
    \frac{1}{\sqrt{\pi}} 
    \Gamma\bra{\frac{\sqrt{1+4m}}{2}}
    \bra{\frac{|\omega| r_{*}^{\rm fin}}{2}}^{-\frac{1}{2}(\sqrt{1+4m}-1)}.
\end{align}
Therefore, modes with smaller $|\omega|$ have a larger growth rate as long as $m>0$.\footnote{For sufficiently small $|\omega|$ with $|\omega| r_{*}^{\rm ini}\ll 1$, the growing mode~$\psi_\omega^{\rm grow}$ is well approximated by the first term in the right-hand side of \eqref{eq:seriesRstar0} throughout $r_{*}^{\rm fin} \le  r_{*} \le r_{*}^{\rm ini}$, which is not oscillating.
Hence, the growth rate~$|\psi_{\omega}^{\rm grow}(r_{*}^{\rm fin})/\psi_{\omega}^{\rm grow}(r_{*}^{\rm ini})|$ does not depend on $\omega$.
On the other hand, for sufficiently large $|\omega|$ with $|\omega| r_{*}^{\rm fin}\gg 1$, $\psi_{\omega}^{\rm grow}$ is oscillating throughout $r_{*}^{\rm fin} \le  r_{*} \le r_{*}^{\rm ini}$ and does not grow.
}

Alternatively, one can regard the system~\eqref{eq:waveeq_toy} as the Klein-Gordon system with the potential~$V_{\rm toy}$ in two dimensions.
For the Klein-Gordon system, one can define the energy in the standard way.
Since the energy is not oscillating, one can directly compare the energy, and obtain the growth rate.
This estimation leads us to the same conclusion as above.

Having clarified the nature of the growing mode in the toy model~\eqref{eq:waveeq_toy}, let us discuss the growing mode for the monopole perturbations in the DHOST theory.
For the monopole perturbations in the DHOST theory, 
we have $m=3/4$ [see Eq.~\eqref{eq:Vtildeexp}], and hence $\psi_\omega$ diverges in the limit~$r_{*}\to 0$. 
Since we have identified $r_{*}^{\mathrm{cut}}=r_{\mathrm{min}}$ and $r_{\mathrm{min}}\sim 3.54\,r_{\mathrm{g}}$, smaller $r_{*}^{\mathrm{cut}}$ corresponds to smaller $r_{\mathrm{g}}$. 
Therefore, for smaller $r_{\mathrm{g}}$, the monopole perturbation in the DHOST theory would become larger.

It is necessary to mention an issue regarding the initial surface. 
Here, we investigate the field evolution in the toy model imposing the initial conditions on $r_{*}=\mathrm{const.}$~surface. 
However, the surface is not physically appropriate for the initial surface of the Cauchy problem. 
Nevertheless, we expect that we can understand the qualitative characters of the growing phase with the analysis of the toy model.

\section{Static monopole perturbation}
\label{app:staticmode}

In the main text, we explore the dynamical evolution of the monopole perturbations.
As a complementary analysis, in this Appendix, we investigate the static mode of the monopole perturbations in the DHOST theory. 
Let us consider the solution~$\psi_{0}(r_{*})$ to the zero-energy Schr\"odinger equation, 
\begin{align}
    \frac{\mathrm{d}^{2}\psi_{0}}{\mathrm{d}r_{*}^{2}} - V_{\mathrm{eff}} \psi_{0} = 0.
    \label{eq:zeroenergyeq}
\end{align}
The general solution to Eq.~\eqref{eq:zeroenergyeq} is given by
\begin{align}
    \psi_{0}(r) = 
    \frac{c_{1}}{r^{1/4}}
    + \frac{c_{2}}{r^{1/4}} 
    \brb{
      r + \frac{r_{\mathrm{g}}}{2} \ln \norm{ \frac{r - r_{\mathrm{g}}}{r+r_{\mathrm{g}}} }
    },
    \label{eq:riccatiphi}
\end{align}
where $c_{1}$ and $c_{2}$ are integration constants.\footnote{Note that $c_{1}$ and $c_{2}$ have dimensions of (length)$^{1/4}$ and (length)$^{-3/4}$, respectively. } 
We can construct the static mode of the metric perturbation from $\psi_{0}$.\footnote{
If $\tilde t$ were the physical time, one could analyze the stability of the system as follows.  
One way is to use the nodal theorem~\cite{messiah2014quantum}. 
The absence of node corresponds to the absence of eigenstate with negative energy, which implies that the system is stable. 
The solution~$\psi_{0}$ with $c_{2}=0$ satisfies the boundary conditions~$\psi_{0}|_{r_{*}\to-\infty} = \mathrm{const.}$ and $\psi_{0}|_{r_{*}\to\infty}=0$, and it does not have any nodes. 
This implies that, if $\tilde{t}$ is the physical time, the system is marginally stable. 
Another way to analyze the stability is the ${\cal S}$-deformation method~\cite{Kodama:2003jz,Ishibashi:2003ap,Kodama:2003kk,Kimura:2017uor,Kimura:2018eiv}. 
We note that a function~${\cal S}$ used in the ${\cal S}$-deformation method can be constructed from the logarithmic derivative of $\psi_{0}$ with $c_{2}=0$. 
However, since $\tilde t$ is not the physical time in reality, in the main text we study the initial value problem and check the dynamical stability explicitly.
}
The components of the static mode of the metric perturbations~$H_{0}$, $H_{1}$, and $H_{2}$ in Eq.~\eqref{eq:perturbationmetric} are given by
\begin{align}
\label{eq:staticHs}
\begin{split}
    H_{0} &= 
    \frac{c_{2} r_{\mathrm{g}}^{3/2} r_{\mathrm{s}}^{1/4}(2D_{1}+3\MPl^{2})}{2 \sqrt{D_{1}}\MPl^{2}(r^{2}-r_{\mathrm{g}}^{2})},
    \\
 H_{1} &=
    \frac{c_{1}r_{\mathrm{g}}^{1/4}}{r\sqrt{D_{1}r_{\mathrm{g}}}}
    -\frac{c_{2}r_{\rm s}^{1/4}}{r\sqrt{D_1r_{\rm g}}}
    \brc{
     \frac{D_1(r^{3}+r_{\mathrm{g}}^{2} r_{\mathrm{s}})}{2
    \MPl^{2}
    (r^{2}-r_{\mathrm{g}}^{2})
     }
     +\frac{r_{\rm g}}{2}
     \brb{2r_{\rm g}r-\frac{r_{\rm g}r_{\rm s}(r^2+3r_{\rm g}^2)}{4(r^2-r_{\rm g}^2)}-(r^2-r_{\rm g}^2) \ln \norm{
       \frac{r-r_{\mathrm{g}}}{r+r_{\mathrm{g}}}
      }}
    },
    \\
    H_{2} &=
    - \frac{c_{2} r_{\mathrm{s}}^{1/4}r^{2}[\MPl^{2} r_{\mathrm{g}}^{2}-D_{1}(r^{2} - r_{\mathrm{g}}^{2})]}{\MPl^{2}\sqrt{D_{1}r_{\mathrm{g}}}(r^{2} - r_{\mathrm{g}}^{2})^{2}}.
\end{split}
\end{align}

We explore the characters of the static mode of the metric perturbations. 
First, we clarify the regularity at the Schwarzschild radius~$r=r_{\rm s}$. 
To do this, we move to the ingoing Eddington-Finkelstein coordinates~$\{v, r, \theta, \varphi\}$ from the Lema\^itre coordinates~$\{\tau, \rho, \theta, \varphi\}$. 
The coordinate transformation law is given by
\begin{align}
    \mathrm{d}\tau 
    = 
    \mathrm{d}v - \frac{1-\sqrt{1-A}}{A} \mathrm{d}r, 
    \qquad
    \mathrm{d}\rho 
    = 
    \mathrm{d}v
    + 
    \frac{1-\sqrt{1-A}}{A\sqrt{1-A}} \mathrm{d}r,
\end{align}
where $A=1-r_{\rm s}/r$.
Then, the background metric and the metric perturbation become
\begin{align}
    \bar{g}_{\mu \nu} \mathrm{d}x^{\mu} \mathrm{d}x^{\nu} 
    &= 
    - A\mathrm{d}v^{2}
    + 2 \mathrm{d}v \mathrm{d}r
    + r^{2} \gamma_{ab}\mathrm{d}x^{a}\mathrm{d}x^{b},
    \\
    h_{\mu \nu} \mathrm{d}x^{\mu} \mathrm{d}x^{\nu} 
    &= 
    h_{vv} \mathrm{d}v^{2}
    + 
    2h_{vr} \mathrm{d}v \mathrm{d}r
    + 
    h_{rr} \mathrm{d}r^{2},
\end{align}
where 
\begin{align}
\begin{split}
    h_{vv} &= H_{0} + 2H_{1} + (1-A)H_{2},
    \\
    h_{vr} &= 
      \frac{(\sqrt{1-A}-1)}{A}H_{0}
    - \frac{2\sqrt{1-A}+A-2}{A\sqrt{1-A}}H_{1}
    + \frac{\sqrt{1-A}-(1-A)}{A}H_{2},
    \\
    h_{rr} &= 
      \frac{1}{2\sqrt{1-A}+2-A} H_{0}
    + \frac{2(\sqrt{1-A}+A-2)}{A^{2}\sqrt{1-A}}H_{1}
    + \frac{1}{2\sqrt{1-A}+2-A}H_{2},
\end{split}
\end{align}
respectively. 
We note that all the components of the metric perturbations are regular at $r=r_{\mathrm{s}}$. 
Therefore, the spacetime described by the perturbed metric is regular at $r=r_{\mathrm{s}}$. 
Let us then see the behaviors of the metric perturbation at the monopole horizon~$r=r_{\mathrm{g}}$. 
To this end, we expand each component of the metric perturbations around $r=r_{\mathrm{g}}$. 
For $r_{\mathrm{s}} \neq r_{\mathrm{g}}$, we have 
\begin{align}
\begin{split}
    h_{vv} &\sim 
    \frac{2c_{1}r_{\mathrm{s}}^{1/4}}{r_{\mathrm{g}}^{3/2}\sqrt{D_{1}}}
    - \frac{c_{2}r_{\mathrm{s}}^{1/4}(2r_{\mathrm{g}}+3r_{\mathrm{s}})}{8\sqrt{D_{1}r_{\mathrm{g}}}(r-r_{\mathrm{g}})},
    \\
    h_{vr} &\sim
    \frac{c_{1}}{r_{\mathrm{g}}r_{\mathrm{s}}^{1/4}\sqrt{D_{1}}} \frac{\sqrt{r_{\mathrm{g}}}-\sqrt{r_{\mathrm{s}}}}{\sqrt{r_{\mathrm{g}}}+\sqrt{r_{\mathrm{s}}}}
    - \frac{c_{2}r_{\mathrm{g}}r_{\mathrm{s}}^{3/4}}{8\sqrt{D_{1}}(r-r_{\mathrm{g}})^{2}},
    \\
    h_{rr} &\sim
    -\frac{2c_{1}}{\sqrt{D_{1}}r_{\mathrm{s}}^{1/4}(\sqrt{r_{\mathrm{g}}}+\sqrt{r_{\mathrm{s}}})^{2}}
    - \frac{c_{2}r_{\mathrm{g}}^{2}r_{\mathrm{s}}^{1/4}}{4 \sqrt{D_{1}}(\sqrt{r_{\mathrm{g}}}+\sqrt{r_{\mathrm{s}}})(r-r_{\mathrm{g}})^{2}}.
\end{split}
\end{align}
On the other hand, for $r_{\mathrm{s}}=r_{\mathrm{g}}$, each component is expanded as
\begin{align}
\begin{split}
    h_{vv} &\sim 
    \frac{2c_{1}}{r_{\mathrm{g}}^{5/4}\sqrt{D_{1}}}
    - \frac{5 c_{2} r_{\mathrm{g}}^{3/4}}{8 \sqrt{D_{1}}(r-r_{\mathrm{g}})},
    \\
    h_{vr} &\sim
     \frac{c_{1}(r-r_{\mathrm{g}})}{4 r_{\mathrm{g}}^{9/4}\sqrt{D_{1}}}
    - \frac{c_{2}r_{\mathrm{g}}^{7/4}}{8\sqrt{D_{1}}(r-r_{\mathrm{g}})^{2}},
    \\
    h_{rr} &\sim
    - \frac{c_{1}}{2 r_{\mathrm{g}}^{5/4}\sqrt{D_{1}}}
    -\frac{c_{2}r_{\mathrm{g}}^{7/4}}{8\sqrt{D_{1}}(r-r_{\mathrm{g}})^{2}}.
\end{split}
\end{align}
For both cases, 
each component of the metric perturbations is singular at $r=r_{\mathrm{g}}$ as long as $c_{2} \neq 0$. 
Furthermore, in order to clarify the regularity of the perturbed spacetime, we evaluate the Misner-Sharp mass. 
For the $r_{\mathrm{g}} \neq r_{\mathrm{s}}$ case, the Misner-Sharp mass around $r=\infty$ and $r=r_{\mathrm{g}}$ satisfies
\begin{align}
    G M_{\mathrm{MS}} \bigl|_{r \to \infty} \sim 
    \frac{r_{\mathrm{s}}}{2} 
    + \epsilon
      \bra{
      \frac{c_{1}r_{\mathrm{s}}^{1/4}}{\sqrt{D_{1}r_{\mathrm{g}}}}
      - \frac{5c_{2} r_{\mathrm{g}}^{3/2} r_{\mathrm{s}}^{1/4}}{2 \sqrt{D_{1}} r}
      }, 
    \qquad
    G M_{\mathrm{MS}} \bigl|_{r \to r_{\mathrm{g}}} \sim 
    \frac{r_{\mathrm{s}}}{2}
    + \epsilon
    \brb{
     \frac{c_{1}r_{\mathrm{s}}^{1/4}}{\sqrt{D_{1}r_{\mathrm{g}}}}
     - 
     \frac{c_{2}r_{\mathrm{g}}^{3/2}r_{\mathrm{s}}^{1/4}(r_{\mathrm{g}}-r_{\mathrm{s}})}{8 \sqrt{D_{1}}(r-r_{\mathrm{g}})^{2}}
    },
\end{align}
respectively, while for the $r_{\mathrm{g}} = r_{\mathrm{s}}$ case, these expansions reduce to
\begin{align}
    G M_{\mathrm{MS}}\bigl|_{r \to \infty} \sim 
    \frac{r_{\mathrm{s}}}{2} 
    + \epsilon
      \bra{
      \frac{c_{1}}{r_{\mathrm{s}}^{1/4}\sqrt{D_{1}}}
      - \frac{5c_{2} r_{\mathrm{s}}^{7/4}}{2 \sqrt{D_{1}} r}
      },
      \qquad
      G M_{\mathrm{MS}} \bigl|_{r \to r_{\mathrm{s}}} \sim 
      \frac{r_{\mathrm{s}}}{2} 
    + \epsilon
      \brb{
      \frac{c_{1}}{r_{\mathrm{s}}^{1/4}\sqrt{D_{1}}}
      - 
      \frac{7c_{2} r_{\mathrm{s}}^{7/4}}{16 \sqrt{D_{1}}(r-r_{\mathrm{s}})}
      },
\end{align}
respectively. 
The Misner-Sharp mass diverges at $r=r_{\mathrm{g}}$ as long as $c_{2} \neq 0$. 
Moreover, the Kretschmann invariant~$R_{\mu \nu \rho \lambda} R^{\mu \nu \rho \lambda}$ also diverges at $r=r_{\mathrm{g}}$. 
Therefore, we conclude that the perturbed spacetime with monopole hair is singular at $r=r_{\mathrm{g}}$ as long as $c_{2} \neq 0$.

Next, we discuss the behavior of the metric perturbations in the asymptotic flat region. 
To this end, we transform the metric into a diagonal form by performing the following coordinate transformation:
\begin{align}
    \mathrm{d}v = \sqrt{k_{1}} \mathrm{d}T + k_{2}(r) \mathrm{d}r,
\end{align}
where 
\begin{align}
    k_{1} = 1 - \epsilon \frac{c_{2}r_{\mathrm{s}}^{1/4}}{\MPl^{2}} \sqrt{\frac{D_{1}}{r_{\mathrm{g}}}},
    \qquad{}
    k_{2}(r) = 
      \frac{1}{A} 
    + \epsilon
      \bra{
         \frac{\sqrt{1-A}}{A^{2}}H_{0}
       + \frac{2-A}{A^{2}\sqrt{1-A}}H_{1}
       + \frac{\sqrt{1-A}}{A^{2}}H_{2}
      }.
\end{align}
Then, expanding the perturbed metric~$g_{\mu \nu}=\bar g_{\mu \nu} + \epsilon h_{\mu \nu}$ in the asymptotic flat region, we have
\begin{align}
    g_{\mu \nu} \mathrm{d}x^{\mu} \mathrm{d}x^{\nu}
    \sim 
    - \mathrm{d}T^{2} + \mathrm{d}r^{2} + r^{2} \gamma_{ab}\mathrm{d}x^{a}\mathrm{d}x^{b}
    + 
    \frac{r_{\mathrm{s}}}{r}\bra{
          1
        + \epsilon \frac{2 c_{1} }{r_{\mathrm{s}}^{3/4}\sqrt{D_{1}r_{\mathrm{g}}}}
      }
      (\mathrm{d}T^{2} + \mathrm{d}r^{2}).
\end{align}
Therefore, the perturbed metric reduces to the Minkowski metric up to ${\cal O}(r^{-1})$. 
In addition, we can easily check $\partial_{r} g _{\mu \nu}={\cal O}(r^{-2})$ and $\partial_{r}^{2} g _{\mu \nu}={\cal O}(r^{-3})$ in the asymptotic flat region. 
From these observations, the spacetime described by the perturbed metric is an asymptotically flat spacetime.

Finally, we investigate the spacetime with $c_{2}=0$. 
In this case, the components of the metric perturbations are given by
\begin{align}
    H_{0}=0, \qquad
    H_{1}=\frac{c_{1}r_{\mathrm{s}}^{1/4}}{r\sqrt{D_{1}r_{\mathrm{g}}}}, \qquad
    H_{2}=0.
\end{align}
As a result, the perturbed metric in the diagonal form is given by
\begin{align}
    g_{\mu \nu}\mathrm{d}x^{\mu} \mathrm{d}x^{\nu} = 
    -\bra{1 - \frac{r_{\mathrm{s}}}{r}
     -\epsilon \frac{2 c_{1} r_{\mathrm{s}}^{1/4}}{r \sqrt{D_{1} r_{\mathrm{g}}}}}\mathrm{d}T^{2}
    + \bra{ 1 - \frac{r_{\mathrm{s}}}{r}
    - \epsilon \frac{2 c_{1} r_{\mathrm{s}}^{1/4}}{r \sqrt{D_{1}r_{\mathrm{g}}}} }^{-1} \mathrm{d}r^{2}
    + r^{2} \gamma_{ab}\mathrm{d}x^{a}\mathrm{d}x^{b}.
    \label{eq:metricdiag}
\end{align}
We can read from this metric that the perturbation shifts the Schwarzschild radius as follows:
\begin{align}
    r_{\mathrm{s}} \to \tilde{r}_{\mathrm{s}} \equiv r_{\mathrm{s}} + \epsilon\, \frac{2c_{1}r_{\mathrm{s}}^{1/4}} {\sqrt{D_{1}r_{\mathrm{g}}}}. 
\end{align}
In terms of the coordinate system~$\{T,r,\theta,\varphi\}$, 
the scalar field is written as
\begin{align}
    \phi(T,r)  = 
    q \bra{
    T + \sqrt{\tilde{r}_{\mathrm{s}}r} + \tilde{r}_{\mathrm{s}} \ln \left| \frac{\sqrt{r} - \sqrt{\tilde{r}_{\mathrm{s}}}}{\sqrt{r} + \sqrt{\tilde{r}_{\mathrm{s}}}} \right|
    }.
    \label{eq:perturbedscalar}
\end{align}
With Eq.~\eqref{eq:perturbedscalar}, the kinetic term of the scalar field~$X=\phi^{\mu} \phi_{\mu}$ can be calculated to be
\begin{align}
    X = -q^{2} + {\cal O} (\epsilon^{2}).
\end{align}
Thus, the perturbed spacetime described by the metric~\eqref{eq:metricdiag} belongs to the family of stealth Schwarzschild solutions.
In this case, the perturbations just shift the the mass of the black hole while keeping the value of $X$ up to the first order in perturbations.
Such a solution can be regarded as trivial and hence is not considered usually.
Therefore, we can set $c_{1}=0$.

\medskip
Let us summarize the results obtained in this Appendix. 
If we impose the regularity condition at $r=r_{\rm g}$ by setting $c_{2}=0$, the solution of the static perturbation describes the shift of the mass parameter of the Schwarzschild spacetime. 
Otherwise, the general solution of the static perturbation contains monopole scalar hair, and the perturbed metric is singular at $r=r_{\rm g}$. 
We note that our numerical calculation of the time evolution of the monopole perturbations suggests that the perturbed metric
with the monopole hair is not realized in a dynamical formation process, 
because the singular behavior of the monopole perturbations near $r=r_{\rm g}$ did not appear in our numerical calculation.

\bibliographystyle{JHEP}
\bibliography{bibliography}

\end{document}